\documentclass[longauth]{aa}
\usepackage{graphicx}
\usepackage{txfonts}
\usepackage{natbib}
\usepackage{natbib}
\bibpunct{(}{)}{;}{a}{}{,}
\usepackage[normalem]{ulem}

\begin{document}

  \title{The Milky Way bar and bulge\\ revealed by APOGEE and {\it Gaia}  EDR3}

   \author{A. B. A. Queiroz\inst{1,2},
   C. Chiappini\inst{1,2},
   A. Perez-Villegas\inst{3,4},
   A. Khalatyan\inst{1},
   F. Anders\inst{5,2},
   B. Barbuy\inst{3}
   B. X. Santiago\inst{6,2},\\
   M. Steinmetz\inst{1},
   K. Cunha\inst{7,8},
   M. Schultheis\inst{9},
   S. R. Majewski\inst{10},
   I. Minchev\inst{1},
   D. Minniti\inst{11,12},
   R. L. Beaton\inst{13},\\
   R. E. Cohen\inst{14},
   L. N. da Costa\inst{2},
   J. G. Fern\'andez-Trincado\inst{15,16},
   D. A. Garcia-Hern\'andez\inst{17,18},
   D. Geisler\inst{19,20,21},
   S. Hasselquist\inst{22,23},
   R. R. Lane\inst{16},
   C. Nitschelm\inst{24},
   A. Rojas-Arriagada\inst{25,26},
   A. Roman-Lopes\inst{21},
   V. Smith\inst{27}
   G. Zasowski\inst{22}
}

   \authorrunning{A. Queiroz et al.}
   \titlerunning{The Milky Way's bar and bulge\\ revealed by APOGEE and {\it Gaia} EDR3}
    \institute{Leibniz-Institut f\"ur Astrophysik Potsdam (AIP), An der Sternwarte 16, 14482 Potsdam, Germany\\
              \email{aqueiroz@aip.de}
        \and{Laborat\'orio Interinstitucional de e-Astronomia - LIneA, Rua Gal. Jos\'e Cristino 77, Rio de Janeiro, RJ - 20921-400, Brazil}
        \and{Department of Astronomy, Universidade de S\~ao Paulo, S\~ao Paulo 05508-090, Brazil}
        \and{Instituto de Astronom\'ia, Universidad Nacional Aut\'onoma de M\'exico, A. P. 106, C.P. 22800, Ensenada, B. C., M\'exico}
        \and{Institut de Ci\`encies del Cosmos, Universitat de Barcelona (IEEC-UB), Carrer Mart\'i i Franqu\`es 1, 08028 Barcelona, Spain}
     \and{Instituto de F\'\i sica, Universidade Federal do Rio Grande do Sul, Caixa Postal 15051, Porto Alegre, RS - 91501-970, Brazil}
     \and{University of Arizona, Tucson, AZ 85719, USA}
     \and{Observat\'orio Nacional, Sao Crist\'ovao, Rio de Janeiro, Brazil}
     \and{Observatoire de la C\^ote d'Azur, Laboratoire Lagrange, 06304 Nice Cedex 4, France}
     \and{Department of Astronomy, University of Virginia, Charlottesville, VA 22904-4325, USA}
     \and{Dept. of Physical Sciences, Faculty of Exact Sciences, Universidad Andres Bello, Fernandez Concha 700, Santiago, Chile}
     \and{Vatican Observatory, V00120 Vatican City State, Italy}
     \and{The Carnegie Observatories, 813 Santa Barbara Street, Pasadena, CA 91101, USA}
     \and{Space Telescope Science Institute, 3700 San Martin Drive, Baltimore, MD 21218, USA}
    \and{Instituto de Astronom\'ia, Universidad Cat\'olica del Norte, Av. Angamos 0610, Antofagasta, Chile}
     \and{Instituto de Astronom\'ia y Ciencias Planetarias de Atacama, Universidad de Atacama, Copayapu 485, Copiap\'o, Chile}
    \and{Instituto de Astrofisica de Canarias (IAC), E-38205 La Laguna, Tenerife, Spain}
     \and{Universidad de La Laguna (ULL), Departamento de Astrofisica, E-38206 La Laguna, Tenerife, Spain}
    \and{Departamento de Astronom\'ia, Universidad de Concepci\'on, Casilla 160-C, Concepci\'on, Chile}
     \and{Instituto de Investigaci\'on Multidisciplinario en Ciencia y Tecnolog\'ia, Universidad de La Serena. Avenida Ra\'ul Bitr\'an S/N, La Serena, Chile}
     \and{Departamento de F\'isica, Facultad de Ciencias, Universidad de La Serena, Cisternas 1200, La Serena, Chile}
    \and{Department of Physics \& Astronomy, University of Utah, Salt Lake City, UT, 84112, USA}
     \and{NSF Astronomy and Astrophysics Postdoctoral Fellow}
     \and{Centro de Astronom{\'i}a (CITEVA), Universidad de Antofagasta, Avenida Angamos 601, Antofagasta 1270300, Chile}
     \and{Instituto de Astrof\'isica, Pontificia Universidad Cat\'olica de Chile, Av. Vicuna Mackenna 4860, 782-0436 Macul, Santiago, Chile}
     \and{Millennium Institute of Astrophysics, Av. Vicu\~{n}a Mackenna 4860, 782-0436, Macul, Santiago, Chile}
     \and{National Optical Astronomy Observatories, Tucson, AZ 85719, USA}
 }

   \date{Received \today; accepted xx.yy.20zz}
\abstract
{
We investigate the inner regions of the Milky Way using data from APOGEE and {\it Gaia} EDR3. Our inner Galactic sample has more than 26\,500 stars within $|X_{\rm Gal}| <5$ kpc, $|Y_{\rm Gal}| <3.5$ kpc, $|Z_{\rm Gal}| <1$ kpc, and we also carry out the analysis for a foreground-cleaned subsample of 8\,000 stars that is more representative of the bulge--bar populations. These samples allow us to build chemo-dynamical maps of the stellar populations with vastly improved detail.
The inner Galaxy shows an apparent chemical bimodality in key abundance ratios [$\alpha$/Fe], [C/N], and [Mn/O], which  probe different enrichment timescales, suggesting a star formation gap (quenching) between the high- and low-$\alpha$ populations. Using a joint analysis of the distributions of kinematics, metallicities, mean orbital radius, and chemical abundances, we can characterize the different populations coexisting in the innermost regions of the Galaxy  for the first time. The chemo-kinematic data dissected on an eccentricity--$|Z|_{\rm max}$ plane reveal the chemical and kinematic signatures of the bar, the thin inner disc, and an inner thick disc, and a broad metallicity population with large velocity dispersion indicative of a pressure-supported component. The interplay between these different populations is mapped onto the different metallicity distributions seen in the eccentricity--$|Z|_{\rm max}$ diagram consistently with the mean orbital radius and V$_{\phi}$ distributions. A clear metallicity gradient as a function of $|Z|_{\rm max}$ is also found, which is consistent with the spatial overlapping of different populations. Additionally, we find and  chemically and kinematically characterize a group of counter-rotating stars that could be the result of a gas-rich merger event or just the result of clumpy star formation during the earliest phases of the early disc that migrated into the bulge. Finally, based on 6D information, we assign stars a probability value of being on a bar orbit and find that most of the stars with large bar orbit probabilities come from the innermost 3 kpc, with a broad dispersion of metallicity. Even stars with a high probability of belonging to the bar show chemical bimodality in the [$\alpha$/Fe] versus [Fe/H] diagram. This suggests bar trapping to be an efficient mechanism, explaining why stars on bar orbits do not show a significant, distinct chemical abundance
ratio signature. 
}
\maketitle

\section{Introduction}
\label{intro}

The Milky Way bulge region, originally identified as a distinct Galactic component by \citet{Baade1946} and \citet{Stebbins1947}, has traditionally been very challenging to observe, because it is a crowded and extincted region (see \citealt{Madore2016} for a review).
Photometric studies of the Galactic bulge towards low extinction windows suggest that the region is old in general \citep[e.g.,][]{Zoccali2003,Renzini2018,Surot2019,Bernard2018}. A spectroscopic sample of lensed dwarfs in the bulge was found to contain a significant population younger than 5 Gyr \citep{Bensby2017}. Optical spectroscopic surveys of the Milky Way traditionally avoid low Galactic latitudes (|b| $\leq$ 5-10) because of the high levels of extinction, especially towards the inner regions.
\citet{Gonzalez2013} used the VISTA Variables in the Via Lactea survey \citep[VVV][]{Minniti2010} to map the mean metallicity throughout the bulge using near-infrared (NIR) photometry, suggesting the existence of a gradient, with the most metal-rich populations concentrated to the innermost regions \citep{Minniti1995}.

Defining a complete sample of the stellar populations in the inner Galaxy has been a challenge. Available spectroscopic samples are traditionally very patchy and fragmented, especially toward the Galactic bulge where heavy extinction and crowding make this area hard to observe. Therefore, most of the spectroscopic data of the Milky Way bulge and bar were limited to a few low-extinction windows (e.g., Baade's Window), or slightly larger latitudes. 

Since the pioneer works of \citet{Rich1988} and \citet{Minniti1992}, the bulge region has been explored by several spectroscopic surveys, such as BRAVA \citep[][]{Rich2007,Kunder2012}, ARGOS \citep[][]{Ness2012}, GIBS \citep[][]{Zoccali2014}, and GES \citep[e.g.,][]{Rojas-Arriagada2014,Rojas-Arriagada2017}, as well as other smaller samples towards lower extinction windows \citep[see][for a review that summarises our knowledge on the Galactic bulge up to 2018]{Barbuy2018}.
The bulge region was confirmed to be dominated by $\alpha$-enhanced stars \citep{McWilliam1994,Cunha2006, Fulbright2007, Friaca2017}, to have a broad metallicity distribution function \citep[MDF;][]{Rich1988,Gonzalez2015,Ness2016b}, to show cylindrical rotation, which is especially contributed by the more metal-rich stars, and to have an X-shape structure which is the result of a buckling bar  \citep[e.g.,][]{Nataf2010, Mcwilliam2010, Saito2012, Li2012, Wegg2017}. It has also been shown that the oldest bulge populations traced by RR Lyrae or very metal-poor stars do not follow the cylindrical rotation \citep{Dekany2013,Gran2015,Kunder2016, Kunder2020, Arentsen2020}. A mix of stellar populations is detected in the Galactic bulge, inferred by the multi-peaked MDF \citep{Zoccali2008,Johnson2013,Ness2013}, usually associated with different kinematics \citep{Hill2011,Babusiaux2010,Babusiaux2014}; for a review see \citet{Babusiaux2016,Barbuy2018}. It has been suggested that the Galactic bulge harbours a more spheroidal, but still barred, metal-poor (with [Fe/H]$ \sim -$0.5) component formed by alpha-enhanced stars, and a more metal-rich ([Fe/H]$\sim$ 0.3) component that forms a boxy bar \citep{Rojas-Arriagada2014, Zoccali2017}, which can split into more components closer to the midplane \citep[see Table 2 of][for a summary]{Barbuy2018}. 

The  field of Galactic archaeology has been transformed in the last two years, firstly by the advent of the second and third early data release of {\it Gaia} \citep[DR2, EDR3][]{GaiaCollaboration2018, GaiaCollaboration2020}, and secondly by the NIR survey ($H$-band) Apache Point Observatory Galactic Evolution Experiment \citep[APOGEE-2][]{Majewski2017, Abolfathi2018} which is currently being extended to the Southern Hemisphere \citep{Ahumada2020}. In 2019, it finally became possible to probe the innermost regions of the Galaxy, much closer to the Galactic plane, with expanded samples of stars with full 6D phase-space information and detailed chemistry. This has opened the possibility for much more detailed studies of the innermost Galactic regions, extending the mapping of the mix of stellar populations to orbital--chemical space \citep[i.e.][]{GarciaPerez2018, Zasowski2019, Trincado2019b, Rojas-Arriagada2019, Sanders2019b, Queiroz2019}.

\begin{figure}{}
\includegraphics[width=8.5cm]{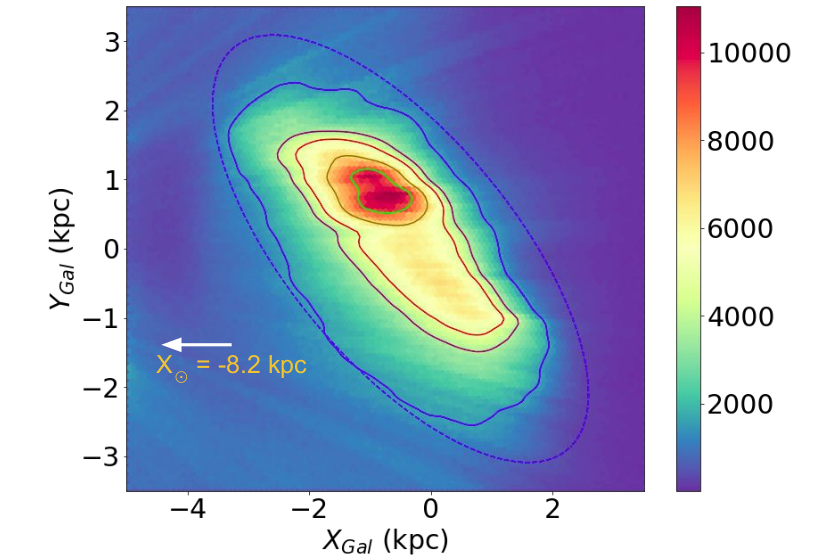}
\caption{Magnified view of the Gaia DR2-derived map of the Galactic bar \citep{Anders2019}. The contours represent the four highest density levels. To guide the eye, an ellipse inclined by 45 deg is drawn in blue. Only RC stars with good {\tt StarHorse} flags close to the Galactic plane ($|Z_{\rm Gal}|<3$ kpc) are shown. The figure contains approximately 30 million stars.}
\label{fig:gdr2bar}
\end{figure}

The latest {\it Gaia} dataset enables the Galactic community to tackle several outstanding questions, regarding for example the shape and kinematics of the Galactic halo \citep[e.g.,][]{Helmi2018, Iorio2019, Myeong2019}, structures in the outer disc \citep[][]{Laporte2020}, the Galactic warp \citep[e.g.,][]{Romero-Gomez2019, Poggio2020,Cheng2020}, the disc spiral structure \citep{Poggio2021}, and also the effect of bar resonances \citep{Kawata2020}. In \citet{Anders2019}, we used the Bayesian {\tt StarHorse} code \citep{Queiroz2018,Santiago2016} to derive photo-astrometric distances and extinctions for around 265 million {\it Gaia} DR2 stars down to magnitude $G<18$. Our calculations allowed the direct detection of the Galactic bar from {\it Gaia} data and stellar density maps  for
the first time. Figure \ref{fig:gdr2bar} shows a zoomed-in version of the the red clump (RC) density map presented by \citet[see][their Fig. 8]{Anders2019}. The breathtaking amount of data (almost 30 million stars with accurate distances and extinctions)  shows the clear shape of an elongated structure around the Galactic centre (GC), associated with the Galactic bar. The map of Figure \ref{fig:gdr2bar} shows the stellar density contours and an ellipse tilted by 45 deg with respect to the Sun--Galactocentric line (Sun--GC), adjusted by eye. This bar orientation is considerably greater than the $\sim$ 30 degrees inferred by most other works \citep[e.g.][]{Wegg2013,Cao2013,Rattenbury2007,Minchev2007,Sanders2019a},  but is in the range of predictions from modelling of the  velocity field of the solar neighbourhood \citep[e.g.][]{Dehnen2000, Minchev2010}. The higher density towards positive $Y$ values is an effect of the lower extinction in that area. 

Although a very clear image of the bar can be seen, the {\tt StarHorse} catalogue of \citet{Anders2019} contains certain caveats that render profound exploration and characterisation of the bulge--bar population difficult. Firstly, the map was derived from parallaxes and photometry only, both of which have elevated uncertainties for the Galactic central region. Secondly, for this sample, {\tt StarHorse} was run with a fixed range of possible extinction values ($A_V<4$ mag), which is not a problem for most regions of the Galaxy, but in the central Galactic plane the extinction can be much higher than $4$ mag (e.g. \citealt{Gonzalez2012, Queiroz2019}). To further characterise the bulge--bar populations, we need large samples of stars observed with IR spectroscopy, which is now becoming possible with APOGEE DR16.

In the present work, we use data from APOGEE which provides spectra for thousands of stars, including those at low latitudes where most of the Milky Way stellar mass is concentrated. The main challenge has been to determine precise distances in order to better define bulge samples with which to constrain, in turn, chemodynamical models. Thanks to the availability and improvements of {\it Gaia}  EDR3 parallaxes in the APOGEE footprint, we derived precise distances and extinctions for the APOGEE stars using the {\tt StarHorse} code \citep{Queiroz2019}, achieving individual distance uncertainties of typically 10\% toward the centre of the Galaxy \citep[see also][]{Schultheis2019}. This makes it finally possible to attempt to disentangle the complex mixture of stellar populations coexisting in the inner Galaxy, which is the goal of the present work.

Although the analysis presented in this paper is based on two much
smaller samples  than the one shown in Figure \ref{fig:gdr2bar}, the rich information provided by combining Gaia  EDR3 and APOGEE allows an unprecedented view of the innermost regions of the Milky Way and the first complete analysis of the sample in orbital space. We are now in a position to offer much tighter observational constraints in chemodynamical simulations of the bulge--bar, contributing to clarifying the current debate over whether the Galactic bulge has a dispersion-dominated component resulting from mergers and/or dissipative collapse of gas \citep{Minniti1992,Zoccali2008}, or if its properties can be completely accounted for by secular dynamical processes forming a buckling bar from pure disc evolution \citep{Debattista2017,Buck2019,Fragkoudi2020}. So far, the broad range of available observational signatures seem to suggest a hybrid scenario in which the metal-poor and the metal-rich populations present in the bulge region would accommodate both the dispersion-dominated and secular-dominated scenarios, respectively (see also discussion in Section 4 of \citealt{Barbuy2018}). Recent results and discussions based on different kinematical populations of RR Lyrae \citep{Kunder2020} also found evidence for bimodal distributions, as well as a small fraction of metal-poor stars and bulge globular clusters; see \citealt{Trincado2020}.

The paper is organised as follows. In Section \ref{sec:data} we describe the spectroscopic data. Section \ref{orbits} describes the computation of velocities and orbital parameters. In Section \ref{sec:samples} we describe our sample selection which consists of an inner-region sample (of around  26\,500 stars) and a cleaned sample that avoids the foreground disc (with around 8\,000 stars).  The chemical and dynamical properties of both samples are described in Sections \ref{chemo} (with particular focus on the observed chemical bimodality) and \ref{kine}. In Section \ref{orbital} we dissect the sample into families in the eccentricity--|Z|$_{max}$ plane. The results and their implications are summarised and discussed in Section \ref{concl}.

\section{Data}
\label{sec:data}

The APOGEE survey is building a detailed chemo-dynamical map extending over all components of the Milky Way. Being the first large spectroscopic survey to explicitly target the central Galactic plane \citep{Zasowski2013,Zasowski2017b} thanks to its NIR spectral range (1.5 - 1.7$\mu m; \ H$-band), APOGEE allows us to determine precise line-of-sight velocities, atmospheric parameters, and chemical abundances, even in highly extincted areas.

APOGEE started as one of the Sloan Digital Sky Survey III \citep[SDSS-III]{Eisenstein2011} programs and is continuing as part of SDSS-IV \citep{Blanton2017}. The observations started in 2011 at the SDSS telescope \citep{Gunn2006} with the northern high-resolution, high signal-to-noise ($R\sim$ $22\,500$, $S/N > 100$) APOGEE spectrograph \citep{Wilson2010}. Since 2017, southern observations have been conducted with a twin spectrograph mounted at the du Pont telescope at Las Campanas Observatory \citep{Wilson2019}.

The latest release of APOGEE data, SDSS DR16 \citep{Ahumada2020}, includes observations from the Southern Hemisphere and contains spectral observation for about $450\,000$ sources. Given the DR16 sky coverage and high-quality observations in the Galactic plane, we can study the Galactic bulge and bar both in the chemical and dynamical space with unprecedented completeness.  Besides the data from APOGEE DR16, we also use the incremental DR16 internal data release which has about  $150\,000$ additional stars observed in March 2020.

Spectral information is obtained through the APOGEE Stellar Parameters and Chemical Abundances Pipeline \citep[ASPCAP][]{GarciaPerez2016,Jonsson2020}. This pipeline compares the observations with a large library of synthetic spectra, determining a best chi-squared fit. The first step in the process is to derive stellar atmospheric parameters and overall abundances of C and N alpha-elements. Then, the second step is to derive abundances from fits to windows tuned for each atomic element. Throughout this paper we use [M/H] (obtained in the first step in ASPCAP) as our metallicity. The studied elements in this paper are: [$\alpha$/Fe], [Fe/H], [O/Fe], [Mg/Fe], [Mn/O], [Mn/Fe], [C/N], and, [Al/Fe]. The APOGEE internal data release has a slightly updated data reduction version (r13).
From the APOGEE catalogue, we select only stars with high S/N, {\tt SNREV} $> 50$, and a good spectral fit from the ASPCAP pipeline, {\tt ASPCAP\_CHI2} $< 25$. 

Besides the APOGEE data, to define a bulge--bar sample we need precise distance measurements.
To this end, we use {\tt StarHorse} \citep{Santiago2016,Queiroz2018} - a Bayesian tool capable of deriving distances, extinctions, and other astrophysical parameters based on spectroscopic, astrometric, and photometric information. In \citet{Queiroz2019}, we combined APOGEE DR16 spectroscopy with {\it Gaia} DR2 parallaxes corrected for a systematic $-0.05$ mas shift \citep{GaiaCollaboration2018, Arenou2018, Zinn2019} and photometry from 2MASS \citep{Cutri2003},  PanSTARRS-1 \citep{Chambers2016}, and AllWISE \citep{Cutri2013} to produce spectro-photometric distances, extinctions, effective temperatures, masses, and surface gravities for around  $388\,000$ stars. In \citet{Queiroz2019}, we also make the same calculation for other major spectroscopic surveys, summing a total of 6 million stars with resulting {\tt StarHorse} parameters. \\

For the data used throughout this paper, we follow the same procedure as in \citet{Queiroz2019} and run StarHorse for the APOGEE DR16 internal release + {\it Gaia} EDR3 parallaxes and the same set of photometry. Corrections were applied to parallaxes as recommended by \citet{Lindegren2020}. With {\it Gaia} EDR3, the resulting distance errors are greatly improved. The samples used along this work have distance uncertainties of around 7\%, while previous computations using {\it Gaia} DR2 allowed us uncertainties of around 10\%. However, the main difference is the improvement on proper motions, as we  discuss in the following section.

\begin{figure*}
\centering
\centering
\includegraphics[width=11.8cm]{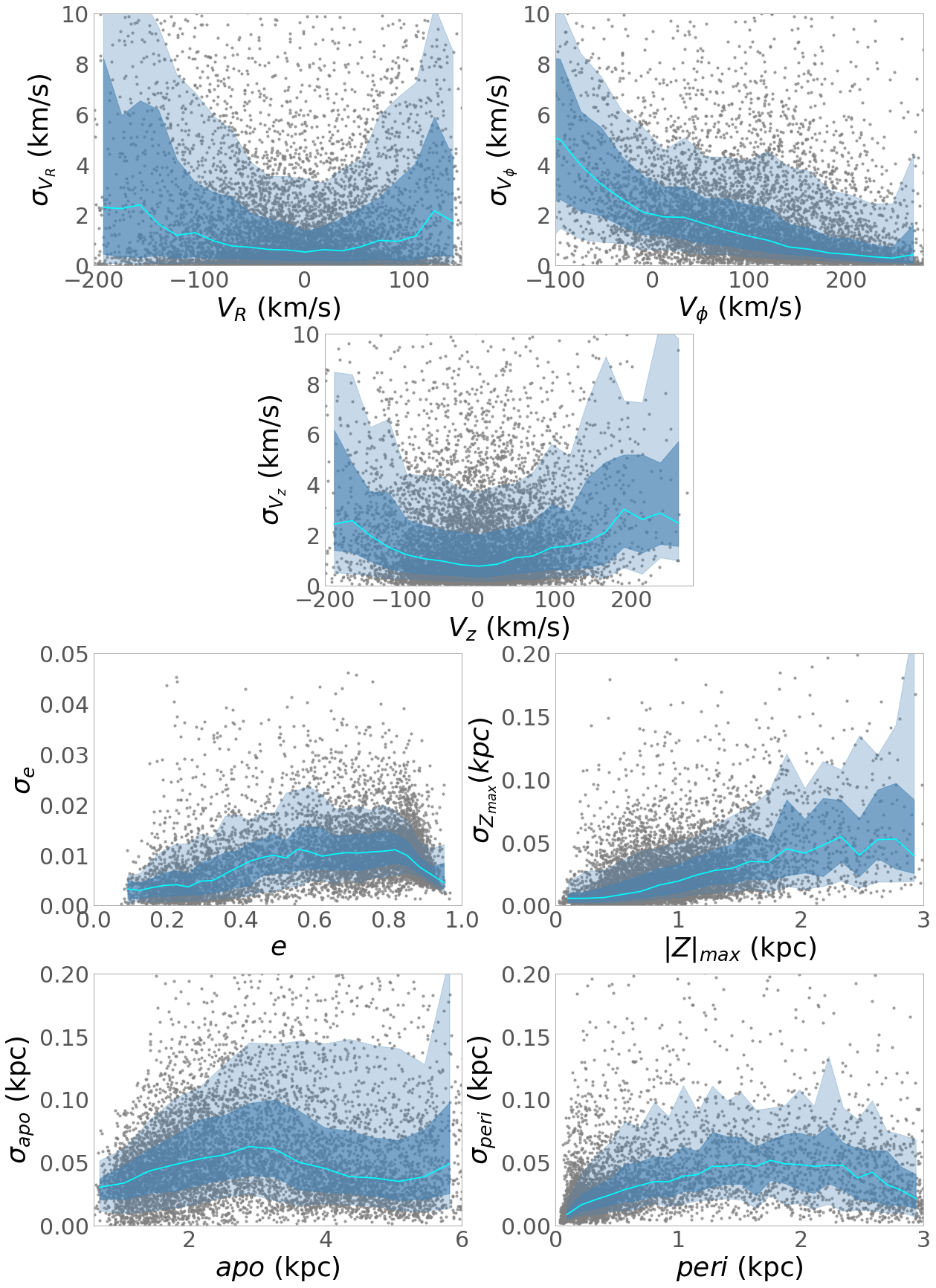}
\caption{Standard error of the cylindrical velocities and orbital parameters. The blue line and shaded areas show the median and standard deviation of 1$\sigma$ and 2$\sigma$  for the distribution.}
\label{fig:orbituncert}
\end{figure*}

\begin{figure}
\centering
\centering
\includegraphics[width=6.5cm]{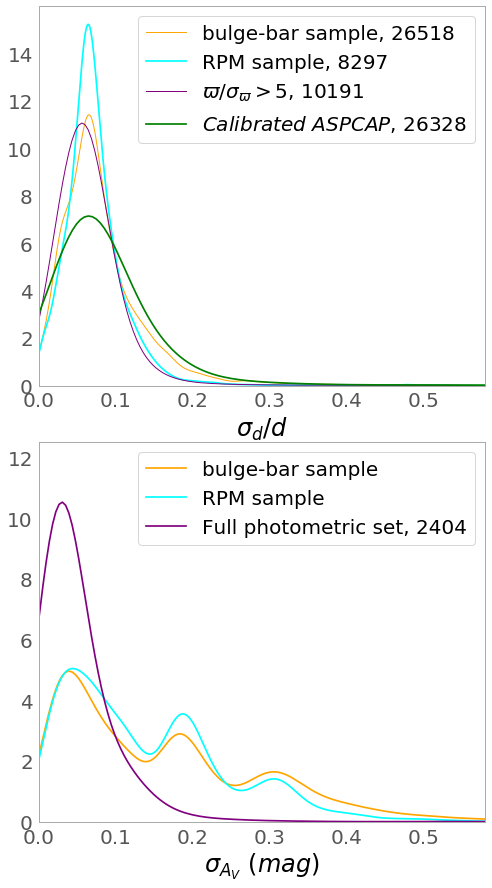}
\caption{Upper panel: Distance uncertainty distributions for the bulge--bar (orange) and  RPM (cyan) samples defined in Sections \ref{sec:barwindow} and \ref{sec:pmselection}, respectively. Also shown are stars with parallax uncertainties smaller than 20\% (magenta) and stars with calibrated ASPCAP parameters (green). Lower panel: Extinction uncertainty distribution for the bulge--bar (orange) and RPM (cyan) samples. Also shown are stars for which all photometric bands are available (magenta). This illustrates that the secondary and tertiary peaks at larger extinction uncertainties seen in our samples are due to stars for which the optical band is not available (see discussion in  \citealt[]{Queiroz2019}).}
\label{fig:baruncert}
\end{figure}

\section{Velocities and orbits}
\label{orbits}

\begin{figure*}{}
\centering
\includegraphics[width=15.5cm]{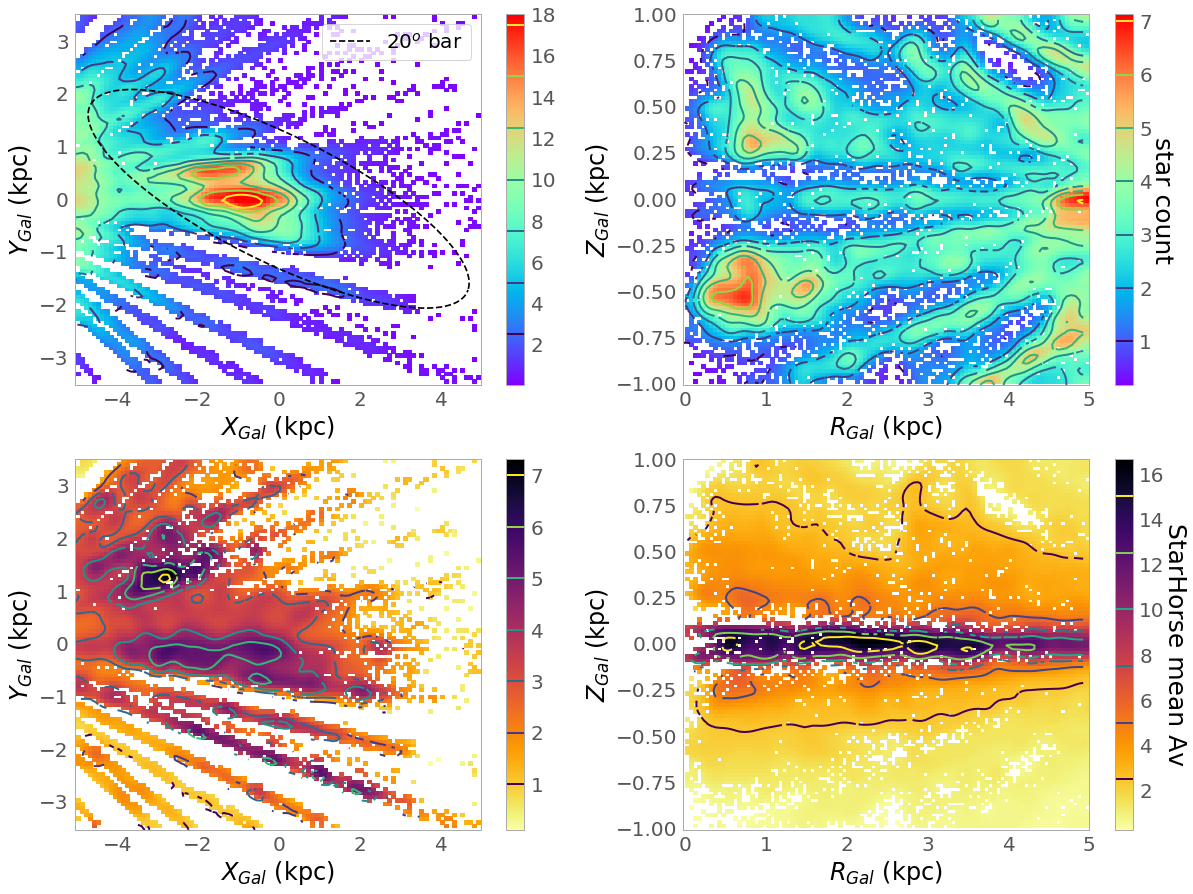}
\caption{Cartesian (left panels) and cylindrical (right panels) projections of the GC using the APOGEE survey and {\tt StarHorse} distances. Upper panels show the map colour-coded by the logarithmic number of stars and lower panels colour-coded by StarHorse extinction. Contours are shown for the densest regions as indicated by the colour bar. An ellipse is drawn in the first panel to indicate the approximate location of the Galactic bar.}
\label{fig:barwindow}
\end{figure*}

\begin{figure*}
\centering
\includegraphics[width=15.5cm]{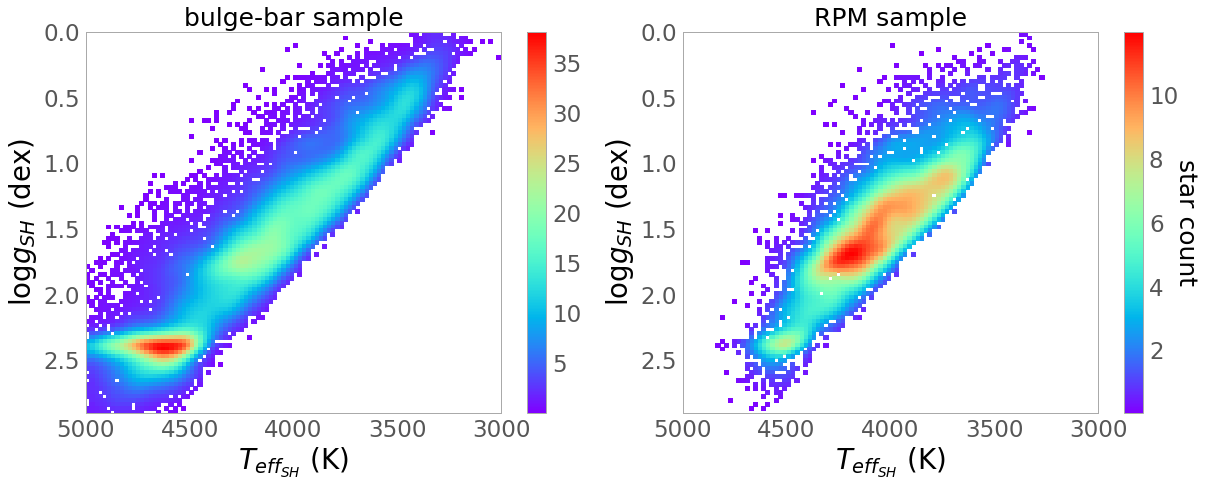}
\caption{Kiel diagrams for the complete bulge--bar sample (left) and the RPM-selected sample (right). This figure illustrates that the innermost regions of the Galaxy are sampled by the more luminous stars. Because more luminous stars tend to be more metal poor, this bias needs to be considered during the analysis.}
\label{fig:barkiel}
\end{figure*}

The combined catalogue APOGEE DR16 internal release + Gaia EDR3 + {\tt StarHorse} gives us access to the 6D phase space of the stars with unprecedented precision. We use the {\it Gaia} EDR3 proper motions, the  line-of-sight velocities (V$_{los}$) measured by APOGEE, and the {\tt StarHorse} distances to calculate the space velocities in Galactocentric cylindrical coordinates. The cylindrical velocity transformations were performed using Astropy library coordinates \citep{astropy2018}, where we use a local standard of rest (LSR) of $v_{\rm LSR}=241$ km/s \citep{Reid2014}, the distance of the Sun to the GC of $R_\odot= 8.2$ kpc, and height of the Sun from the Galactic plane of $Z_\odot=0.0208$ kpc \citep{Bennett2019}. We also note that in all of our diagrams we use the Sun position at X$_{Gal}=-8.2$ kpc. 

We assume the peculiar motion of the Sun with respect to the LSR to be: $(U, V, W)_{\odot}$= (11.1, 12.24, 7.25) km/s \citep{Schoenrich2010}. The resulting components of the velocity we use throughout this paper are the azimuthal velocity, V$_{\phi}$, the radial velocity $V_R$, and the vertical velocity $V_Z$. All these components are with respect to the GC. We also note that $V_R$ $\neq$ $V_{los}$.

As all bodies in the Milky Way move under the Galactic potential, many stars that we find nowadays with a present position at the GC may actually be in a disc or halo orbits. To identify if the stars are from disc, halo, or from bulge--bar components we proceed with the calculation of the orbital parameters. Our Galactic potential includes an exponential disc generated by the superposition of three Miyamoto-Nagai  discs \citep{Miyamoto1975, Smith2015}, a dark matter halo modelled with an NFW density profile \citep{Navarro1997}, and a triaxial Ferrers bar \citep{Ferrers1877,Pfenniger1984}. The total bar mass is $1.2\cdot 10^{10}\, {\rm M}_\odot$, the angle between the bar’s major axis and the Sun--GC line is 25 deg, its pattern speed is 40 km s$^{-1}$ kpc $^{-1}$  \citep{Portail2017,Perez-Villegas2017,Sanders2019b}, and its half-length is 3.5 kpc. To consider the effect of the uncertainties associated with the observational data, we used a Monte Carlo method to generate 50 initial conditions for each star, taking into account the errors on distances, heliocentric line-of-sight velocities, and the absolute proper motion in both components. We integrate those initial conditions forward for 3 Gyr with the NIGO tool \citep{Rossi2015}. From the Monte Carlo experiment, we calculated the median of the orbital parameters for each star: perigalactic distance $R_{\rm peri}$, apogalactic distance $R_{\rm apo}$, the maximum vertical excursion from the Galactic plane |Z|$_{\rm max}$, the eccentricity $e=(R_{\rm apo}-R_{\rm peri})/(R_{\rm apo}+R_{\rm peri})$ and the mean orbital radius, R$_{mean}$ = $(R_{\rm apo}+R_{\rm peri})/2$. In the following sections, we use those orbital parameters when analysing the chemical patterns found in the innermost regions of the Galaxy. We show the uncertainties in the orbital parameters and cylindrical velocities in Figure \ref{fig:orbituncert}. These distributions increase with increasing distance, which is expected because for larger distances we have larger {\tt StarHorse} distance uncertainties. The uncertainties on velocity are larger for retrograde stars (negative v$_\phi$) but are still usually around $5$ km/s. The other components of the velocity show higher uncertainties for faster stars.

One caveat in these calculations is that orbital parameters depend on the model employed. We integrated the orbits in a steady-state gravitational potential. In our model, we do not take into account dynamical friction and the secular evolution of the Galaxy \citep{Hilmi2020}. Also, we do not consider the dynamical effects due to the spiral arms. In Figure \ref{fig:orbtsps} of the Appendix, we show a comparison of the orbital parameters computed using different bar pattern speeds. The comparison gives relative differences of less than 20\% for most of the stars.

\section{Sample selection}\label{sec:samples}

\begin{figure*}
\centering
\includegraphics[width=18.5cm]{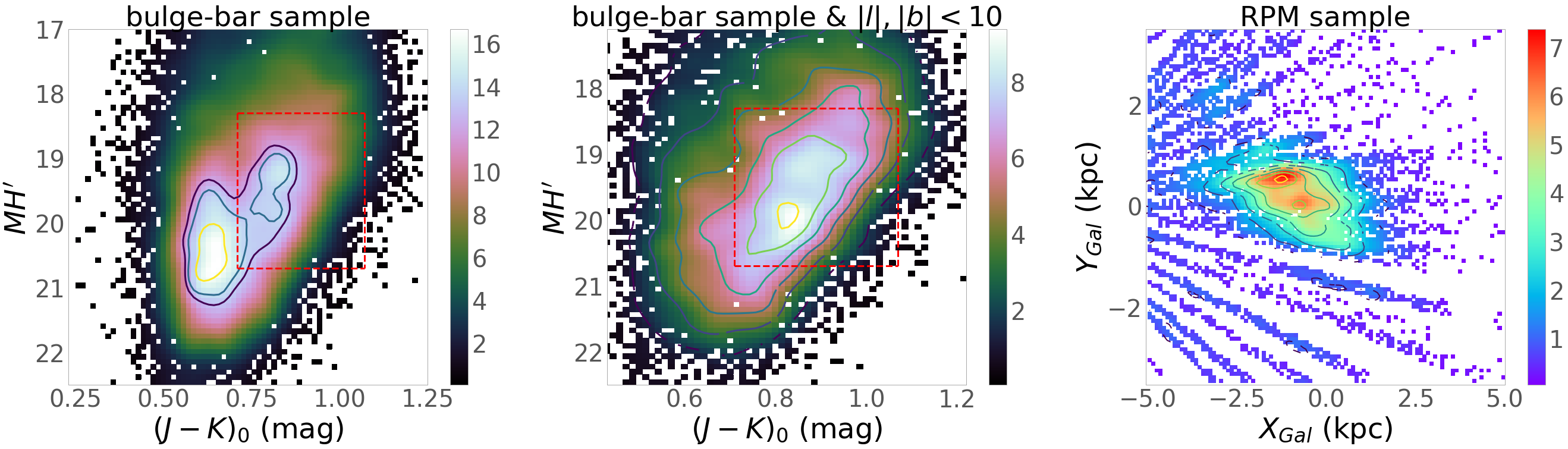}
\caption{Illustration of our RPM selection. Left panel: RPM diagram. Contours show the most dense areas, highlighting two main density groups. Middle panel: Same as left panel, but for the central region  ($|l|,|b|<10$ deg). In both panels, the red dashed box indicates the boundaries of our RPM selection. Right panel: Cartesian density map of stars satisfying the RPM cut.}
\label{fig:pmselection}
\end{figure*}

In this paper we focus our analysis on the inner region of the Milky Way. In particular, we study a window that is symmetric about the GC in all  three dimensions in Galactocentric Cartesian coordinates; see Figure \ref{fig:barwindow} ($|X_{\rm Gal}|<5$ kpc, |$Y_{\rm Gal}|<3.5$ kpc, and $|Z_{\rm Gal}|< 1.0$ kpc).\\
Throughout the paper, we use two samples: (1) the full bulge-bar sample with the geometric cuts (detailed in Sect. \ref{sec:barwindow}), and (2) a cleaned subsample (see Sect. \ref{sec:pmselection}).

The uncertainties on distance and extinction are shown in Figure \ref{fig:baruncert} for the two samples discussed in the following section, the bulge-bar sample and the reduced proper motion sample (RPM). Our stars can be seen to have uncertainties on distance of less than 15$\%$ which would translate to around 1.5 kpc for the stars with the largest errors. The distribution of distance uncertainties shows no big differences with quality cuts such as parallax relative errors $>$ 20$\%$ or using only calibrated ASPCAP inputs. The extinction uncertainties from {\tt StarHorse} has three main peaks at A$_V$ $\sim$ 0.05 mag, A$_V$ $\sim$ 0.2 mag, and A$_V$ $\sim$ 0.3 which are caused by the availability or not of one or more passbands from the full photometric set: \{2MASS, AllWISE, PanSTARRS-1\}. For a further discussion about the uncertainties on these parameters and their correlations please see \citet{Queiroz2018} and \citet{Queiroz2019}.

\subsection{Bulge--bar sample}\label{sec:barwindow}

The full bulge--bar sample has a total of $26\,518$ stars, with typical distance uncertainties of around $7 \%$ (see below). This APOGEE DR16 inner Galactic sample has unprecedented coverage of thousands of stars that reach Galactic latitudes below $|b|<5$. This low latitude range was not covered in previous dedicated surveys such as BRAVA and ARGOS, which were fundamental in revealing the peanut bar shape and in showing the rotation of the stars in the GC \citep[][]{Kunder2012, Ness2013}.
 The density and extinction distributions for the bulge--bar sample can be seen in Figure \ref{fig:barwindow}; the distribution is far less complete in terms of density than Figure \ref{fig:gdr2bar}, but the dense areas in the figure do seem to follow a bar-shaped pattern with higher density around the GC. If we again trace an ellipse by eye around the density contours, we obtain a much smaller inclination angle with respect to the Sun--GC line, of namely around 20 deg, which is much closer to the canonical value of $\sim27$ deg \citep{Bland-Hawthorn2016}. The angle from the ellipse fitted by eye is certainly not precise, but we see that the bar-shaped structure is less inclined than in \citet{Anders2019}.

This seems to confirm the suspicion that the photo-astrometric distances for the bar structure seen in Figure \ref{fig:gdr2bar} are slightly overestimated because the extinction values get saturated at around A$_V =$ 4.  Figure \ref{fig:barwindow} also shows that we still lack data very close to the Galactic plane, $|Z_{\rm Gal}|<$0.2 kpc, as this area remains hidden by very high extinction (e.g. for $|Z_{\rm Gal}|<$0.1 kpc we often observe large-scale extinction $A_V>10$; \citealt{Minniti2014}). The Kiel diagram for this sample is shown in the first panel of Figure \ref{fig:barkiel}, showing that the population in this sample is mainly composed of  red giant branch stars and RC stars.

\subsection{Reduced-proper-motion diagram selection}\label{sec:pmselection}

There are different ways to select a cleaner and more homogeneous bulge--bar sample, avoiding foreground disc stars. Usually, studies of bulge stars select fields in the direction of Baade's window \citep{Babusiaux2010,Hill2011} or fields in the direction of the GC \citep{Zoccali2008,Kunder2012,Rich2012}. We have a massive amount of information about the stars, and in addition to simply selecting the bulge-bar sample we can constrain an even `cleaner' sample. One way to do this is to draw isocontours around the XY density maps. Another way is to look for similarities in the stellar composition. However, we could still be left with disc or halo stars and/or potentially important systematic abundance differences resulting from the fact that stars at different distances will have systematically different luminosities and stellar parameters. An additional abundance pre-selection would bias the study towards the chemical distribution of the bar--bulge components. For our definition of a clean bulge--bar sample, we therefore opt for a selection in  the RPM  diagram. Our goal with this selection  is to clean the most apparent disc contamination without an abrupt cut in distances. 

 The RPM \citep[][]{Faherty2009, Gontcharov2009, Smith2009} is a common tool used to distinguish between distinct kinematical populations. In the RPM diagram, $M_{H'}$ is defined analogously to the absolute magnitude, because the proper motions are also a proxy for the star's distance:
\begin{equation}
M_{H'} = H_{2mass} + 5.0 + \log_{10}(\sqrt{\mu_{\rm RA}^2 + \mu_{\rm DEC}^2}).
\end{equation}

In Figure \ref{fig:pmselection} we show the RPM diagram, $(J-K_s)_0$ versus $M_{H'}$, for the bulge--bar sample defined above. The RPM diagram shows two agglomerations highlighted by the density contour levels, indicating distinct populations \citep[e.g.][]{Holtzman2018}. A cut in  $|l|,|b| <$ 10 (middle panel of Figure \ref{fig:pmselection}) is analogous to a cut selecting the rightmost agglomeration, which is roughly indicated by the red rectangle, showing this cut represents the  innermost population. The left-most agglomeration extends in colour, connecting with the rightmost stellar overdensity. In our selection, the tail of this  population remains because we want to preserve completeness and a more symmetrical colour cut around the rightmost overdensity. The selection of stars inside the red rectangle also results in the exclusion of most of RC  stars, as one can see in Figure \ref{fig:barkiel}.
Our goal with this simple selection is to filter disc stars from our sample with the fewer biases possible to study chemistry and kinematics. We also highlight the fact that the cut in kinematics is minimal; we mostly cut the tails of the proper motion distribution, which have lower density bins. Therefore, the RPM cut is more consistent with a colour cut than a kinematic cut.

With this selection we maintain a relatively homogeneous coverage of the entire inner Galaxy, while removing background and foreground over-densities of disc RC stars.
The RPM diagram selection shown in Fig. \ref{fig:pmselection} results in a more smoothly distributed population around the GC and slightly distorts the density contours found for the purely geometric bulge--bar sample. The squared selection was chosen for simplicity, because the main purpose of this stricter sample is to distinguish whether the results found with the full sample are robust or if they may be significantly biased by the complex mix of stellar populations, the selection function of APOGEE, or systematic errors on  abundance.

\section{Chemical composition}\label{chemo}
\begin{figure*}{}
\centering
\includegraphics[width=15.5cm]{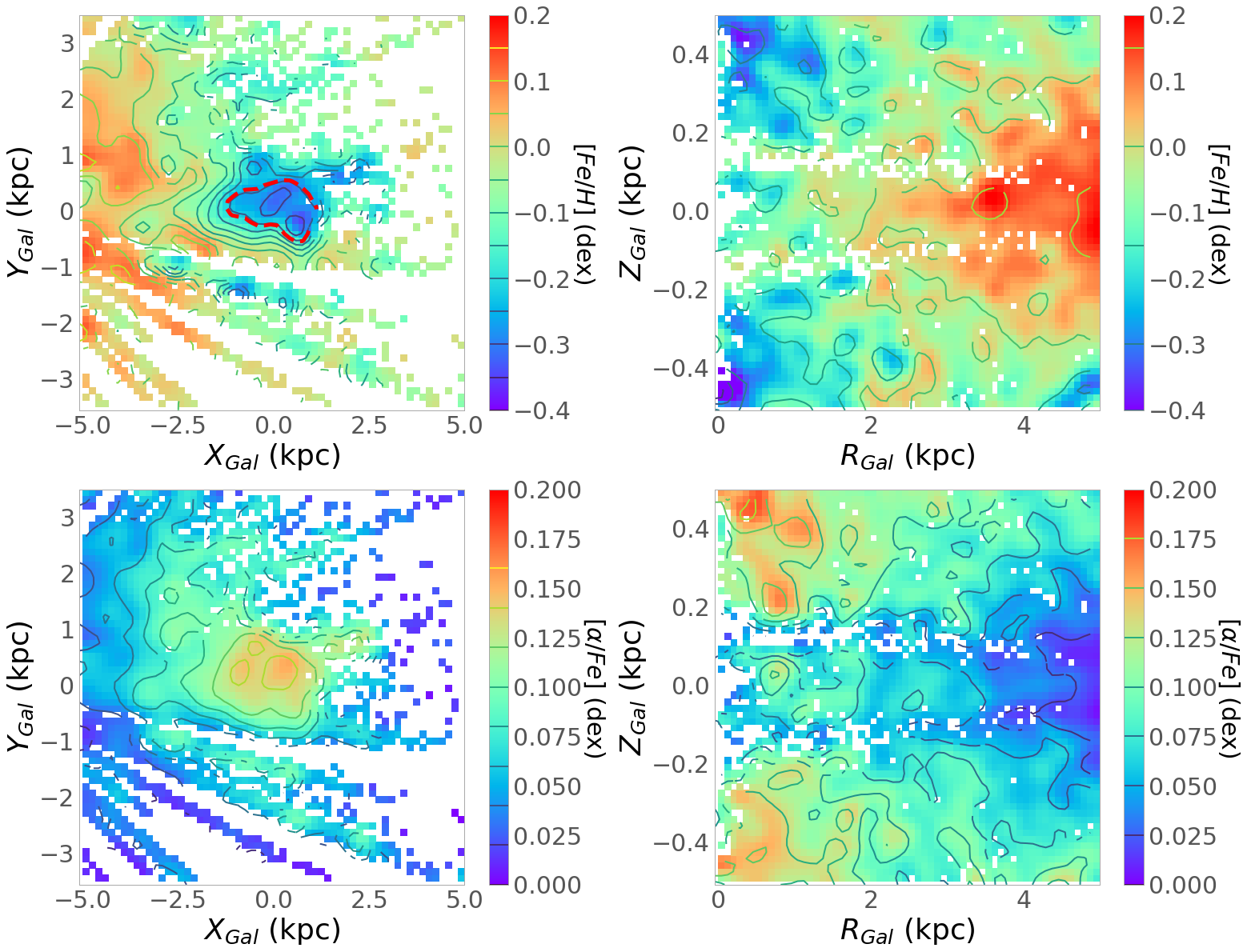}
\caption{ Cartesian (left panels) and cylindrical (right panels) Galactocentric projections of the bulge--bar sample with an extra cut of $|Z|<0.5$ kpc. Bins are colour-coded according to their mean [Fe/H] (upper panels) and [$\alpha$/Fe] (lower panels) content. A red contour is drawn around the metal poor area in the innermost regions of the Milky Way.}
\label{fig:alphapop}
\end{figure*}

\begin{figure*}{}
\centering
\includegraphics[width=15.5cm]{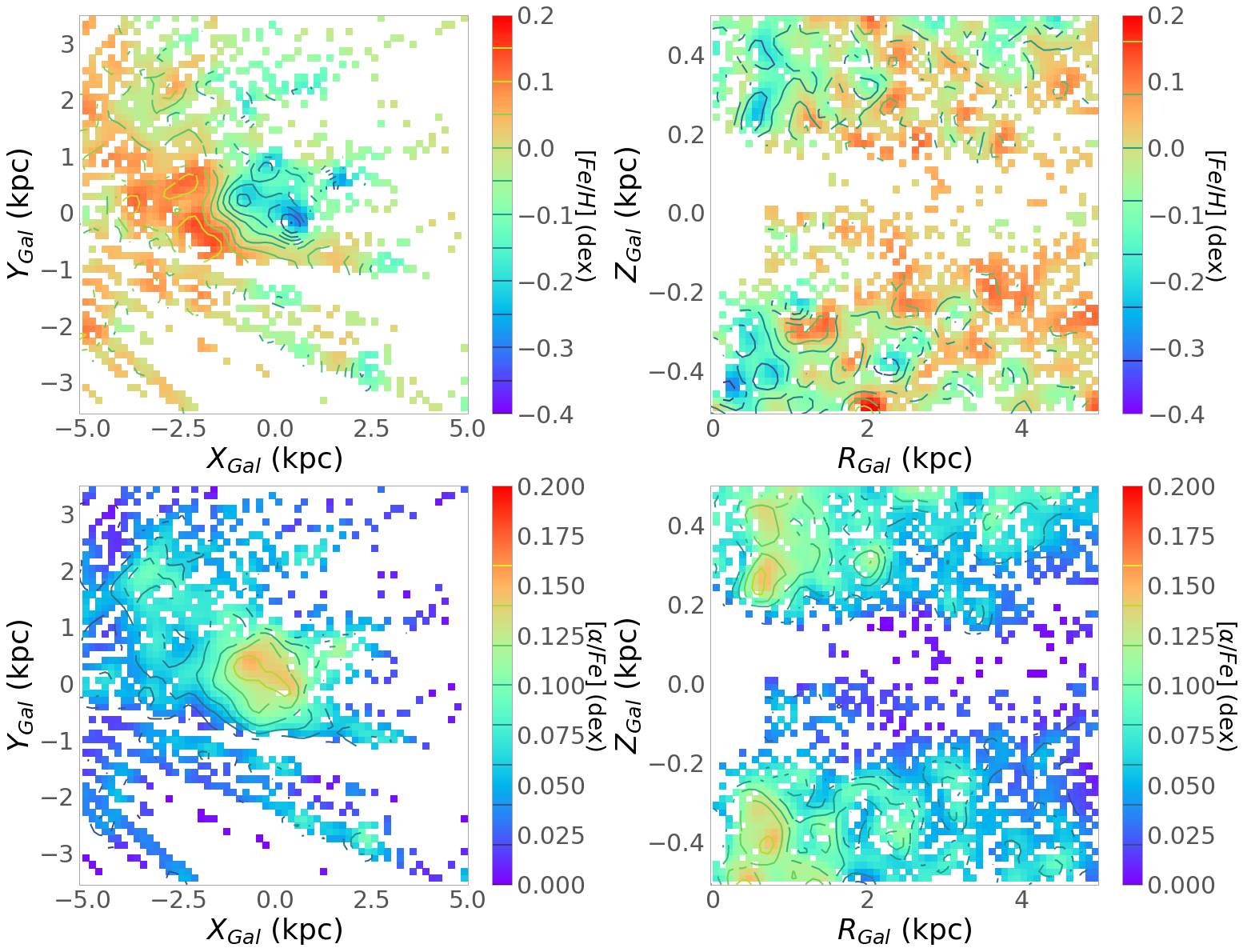}
\caption{As in Fig. \ref{fig:alphapop}, but for the RPM sample, with around  3\,800 stars. We note the lack of stars very close to the midplane, resulting from the fact we do not have {\it Gaia} proper motions for a considerable fraction of these stars.}
\label{fig:alphapopRPM}
\end{figure*}

\subsection{The $\alpha$-elements and metallicity}
\begin{figure}
\includegraphics[width=8.5cm]{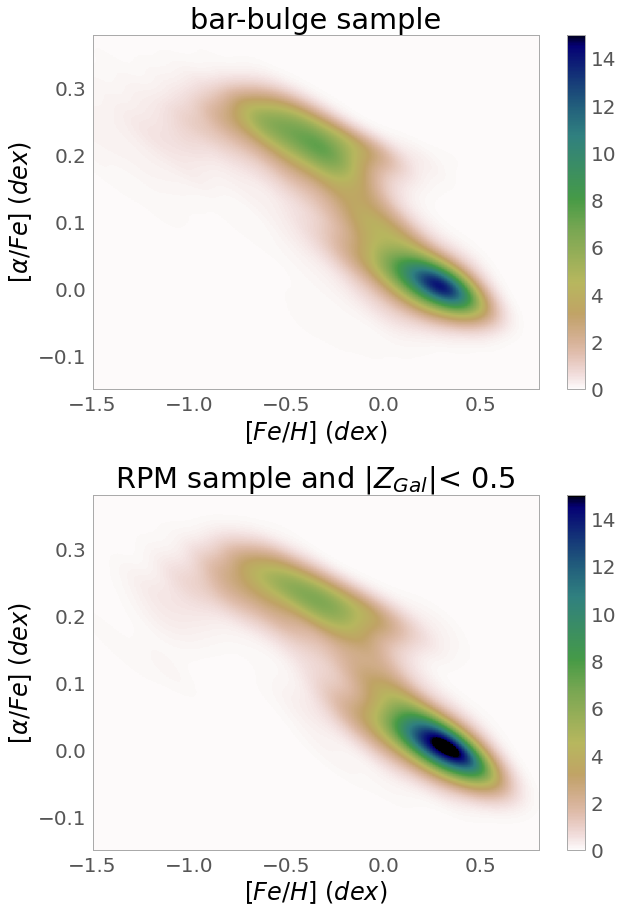}
\caption{ [$\alpha$/Fe] vs. [Fe/H] distributions for the bulge--bar region ($\sim$ 26\,500 stars) and RPM sample ($\sim$ 3\,800 stars with $|Z_{Gal}|<0.5$ kpc), colour-coded according to probability density function. }
\label{fig:alphafe}
\end{figure}

\begin{figure}
\includegraphics[width=8.5cm]{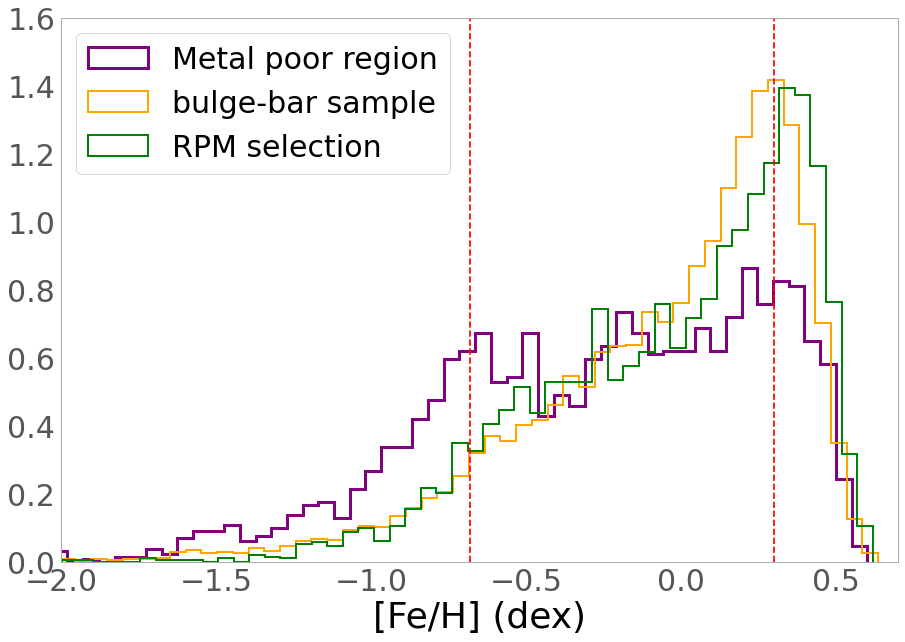}
\caption{MDF for the bulge--bar field, RPM sample, and metal-poor region highlighted in Fig. \ref{fig:alphapop}. The prominent peaks of the distributions are indicated by the vertical dashed lines.}
\label{fig:mdf}
\end{figure}

As mentioned in Section~\ref{intro}, the chemical composition of the bulge--bar region is fairly complex; for example its metallicity distribution has multiple peaks \citep[e.g.,][]{Ness2013, Rojas-Arriagada2014, Rojas-Arriagada2017, Schultheis2017,GarciaPerez2018, Rojas-Arriagada2020}, and the innermost regions of the Milky Way show not only the signature of a bar and a spheroid but also that of the stars from the halo and the thin and thick discs \citep{Minniti1996}.  In particular, it is still debated whether the thin and thick discs might have different chemical signatures in their inner regions  from those of their local counterparts; see discussion in \citep{Barbuy2018,Lian2020}. This is especially the case for the thin disc, as shown by the metallicity gradients with Galactic radius \citep[e.g.][]{Hayden2014, Anders2014, Anders2017a}. Moreover, debris from accreted globular clusters and dwarf galaxies is also expected to populate the central regions of the Milky Way \citep[see][]{Das2020,Horta2020,Trincado2019,Trincado2020b,Trincado2021}.

In this section, we first focus on the main chemical characteristics of our inner Galactic samples as defined in the previous sections. It is important to keep in mind that we have used the ASPCAP [M/H] value as representative of metallicity, as explained in Section \ref{sec:data}. No fundamental difference in results is seen when using [Fe/H] or [M/H] as the proxy for metallicity, but we retain a larger number of metal-rich stars of [M/H]$>$0.2 dex if [M/H] is used (see Section \ref{zmaxecc}).

In the present work we have chosen to focus only on the following four abundance ratios: (a) the classical [$\alpha$/Fe] ratio (as well as [O/Fe] and [Mg/Fe] for consistency checks, although for fewer stars), which is available for the whole sample and is a good tracer of the chemical enrichment timescales \citep[e.g.,][]{Matteucci1991, Haywood2012,  Miglio2021}; (b) [C/N], which is used in the solar vicinity as a cosmic clock \citep{Masseron2015,Martig2016,Hasselquist2019}; and (c) the [Mn/O] and [O/H] ratios which also separate thick and thin disc stars \citep[e.g.][]{McWilliam2013,Barbuy2013,Barbuy2018}.

Figure \ref{fig:alphapop} shows the spatial chemical abundance maps in Cartesian (XY) and cylindrical (RZ) coordinates colour-coded according to  [Fe/H] and [$\alpha$/Fe] abundances for the bulge--bar sample with an extra cut in  Galactic height |Z|$<$0.5 kpc; this sample contains $\sim$ $14\,500$ stars. The map shows an interesting spatial dependency of the metallicity, with a metal-poor ($\alpha$-rich) component that seems to dominate the more central region, a feature that we can now see for the first time in the XY plane. We note that selection effects alone cannot explain this latter structure, because such effects are related to distance, and we can clearly see that the contribution from low-metallicity stars increases towards  the GC, $X_{Gal}$ $\sim$0~kpc, heliocentric distance d $\sim 8$ kpc, and that at greater distances the metallicity starts do increase again (although more data are needed to confirm this point, especially in the Galactic southern hemisphere). In photometric samples of the bulge area as a whole, the metal-rich population seems to dominate, as photometric maps report an increase in the metallicity towards the innermost Galactic regions \citet{Gonzalez2013}. The more detailed data discussed here enable us to see the spatial variations of the mean metallicity for stars closer to the Galactic midplane (0.2 $<$ |Z$_{Gal}$| $<$ 0.5), showing a clear inversion of the radial metallicity gradient in the innermost 1 kpc. In the GC, the metallicity seems to be high again as shown by \citet{Schultheis2019}.  

The RZ projection also shows large metallicity values (and lower [Mg/Fe]) closer to the Galactic midplane, becoming much less prominent at higher latitudes, a result already known from previous studies of the bulge MDF \citep[e.g.,][]{Zoccali2008} inferred in the latitude, longitude space. The projection also shows that the central metal-poor population extends to high Z$_{Gal}$. In the very low Galactic plane, Z$_{Gal}<$0.2 kpc, there is a lack of data due to high extinction \citep[e.g.][]{Minniti2014,Queiroz2019}, and therefore  with the current sample we are not able to  determine whether the innermost population is dominated by metal-rich or metal-poor stars. 

It is beyond the scope of this paper to correct for selection effects, which we plan to do in a future work dedicated to the detailed comparison of our data with chemo-dynamical models. In the case of APOGEE, the lines of sight and magnitude determine the selection function, which can limit the populations in age or chemistry. In an upcoming paper \citep{Queiroz2021} we will use mock simulations to study how these selection effects change our sample. However, the selection function seems to have a minor impact, as illustrated in recent work by \citet{Rojas-Arriagada2020} using APOGEE DR16, and also in work using DR14 \citep{Nandakumar2017}. There appears to be bias towards preferentially observing metal-poor (brighter) objects in the most reddened regions. Here we try to gauge this effect by investigating the RPM sample, which is shown in Figure \ref{fig:alphapopRPM}. This figure shows  a considerable lack of data in the most central regions of the Galaxy at Z$_{Gal}$ $<$ 0.2 kpc compared to the bulge--bar sample. The absence of data in the low Galactic plane in the RPM sample results from the unavailability of {\it Gaia} EDR3  data for the high extinction and crowded areas such as the inner Galaxy. From the bulge--bar sample, around $3\,000$  stars have no {\it Gaia EDR3} proper motions. These are almost all located at low Galactic heights. Given this fact, there is no apparent shift to more metal-poor stars in the central regions sampled by the RPM selection than is seen when analysing the bulge--bar sample. We note that in the inner 200 pc regions, and in particular close to SgrA within the nuclear star cluster, we find a very metal-rich dominant population \citep{Schultheis2019}. In any case, these caveats should be kept in mind when discussing the results that relate chemistry with kinematics and orbital parameters in Sections \ref{kine} and \ref{orbital}, especially in the lower Z$_{Gal}$ regions, and when extracting conclusions from 2D chemical abundance diagrams.

\begin{figure*}{}
\centering
\includegraphics[width=15.5cm]{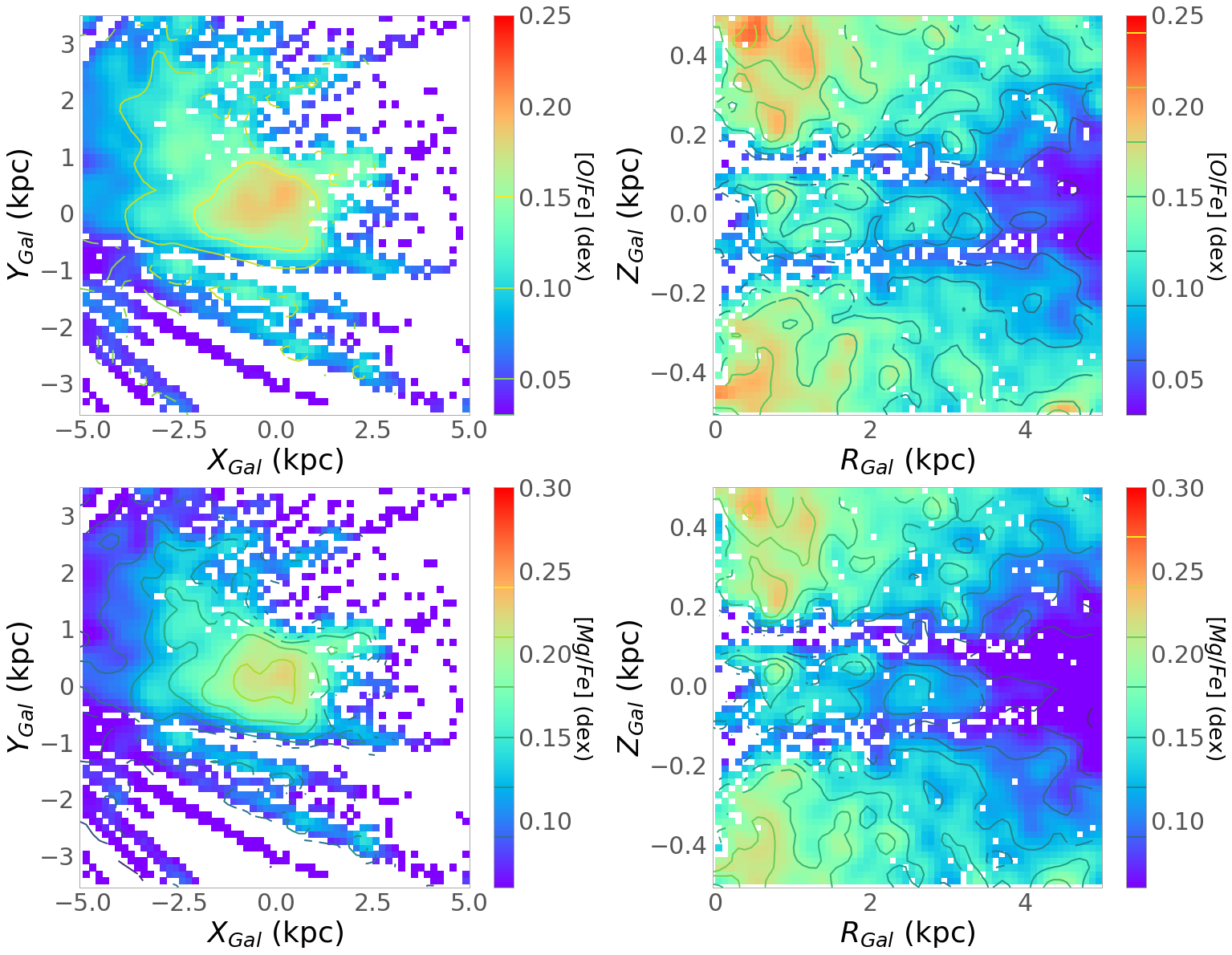}
\caption{Same as Fig. \ref{fig:alphapop} but now bins are colour-coded according to their mean [O/Fe] (upper panels) and [Mg/Fe]  (lower panels). These maps are fully consistent with what was seen before when using the ASPCAP $\alpha$ instead of the individual alpha elements given by the pipeline.}
\label{fig:omgfe}
\end{figure*}

\begin{figure*}{}
\centering
\includegraphics[width=15.5cm]{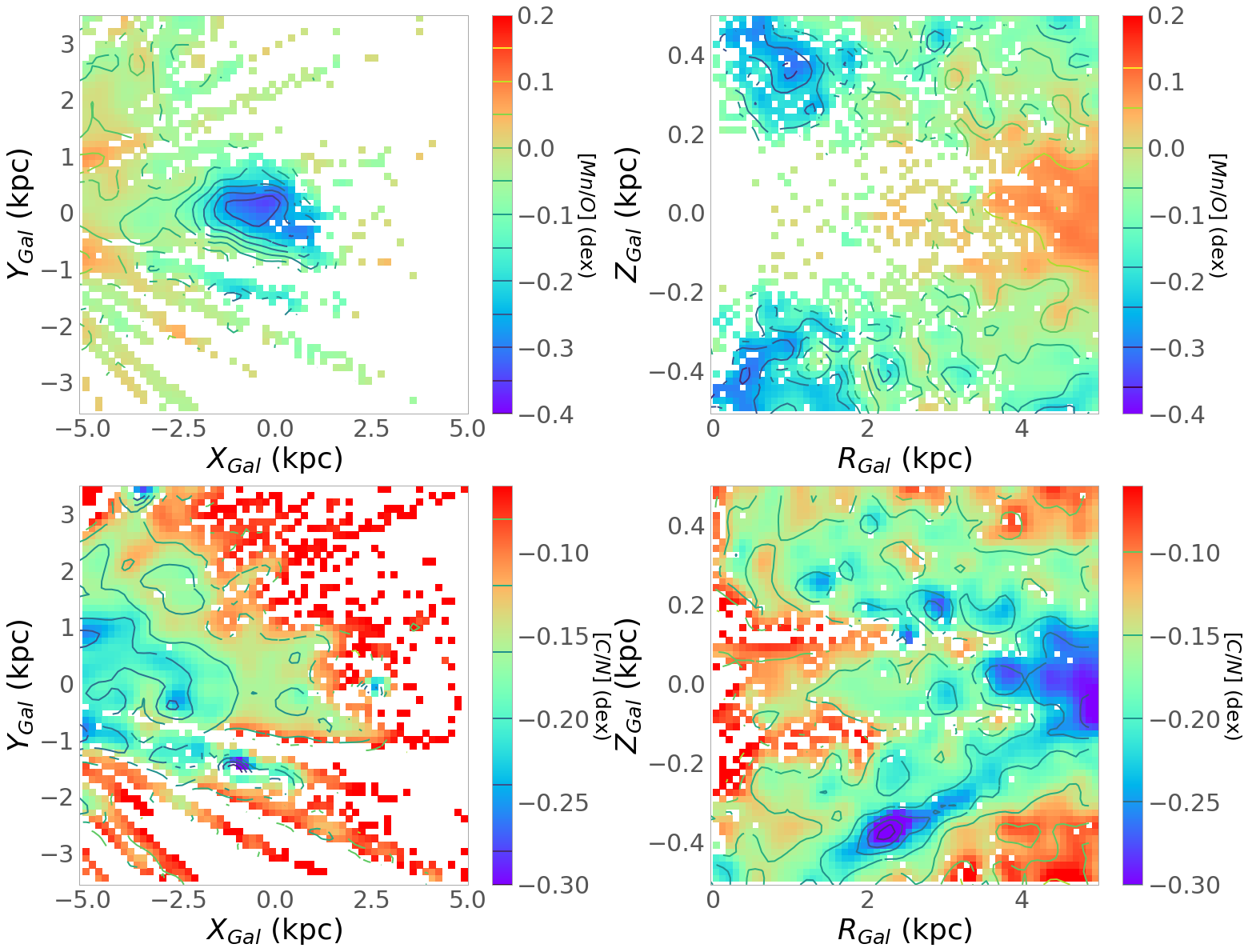}
\caption{Same as Figure \ref{fig:alphapop}, now colour-coded by [Mn/O] (upper panels) and [C/N] (lower panels).}
\label{fig:cnmno}
\end{figure*}

The  [$\alpha$/Fe] versus [Fe/H] plane is now shown for our two samples in Figure \ref{fig:alphafe}. In the figure we use a kernel density estimation from scipy \citep{Virtanen2019} to estimate the probability density function.  In both cases, the sequences show a bimodal distribution with an $\alpha$-rich and $\alpha$-poor populations, with the two subcomponents becoming better defined  when we apply the proper motion selection to remove foreground stars, and confine the sample to near the Galactic midplane. This bimodality was also reported by \citet{Rojas-Arriagada2019} based on APOGEE DR14 data, though in the paper by \citet{Queiroz2019} and here the depression between the two peaks is significantly clearer, with the two sequences markedly separated. 

\begin{figure}
\includegraphics[width=8.5cm]{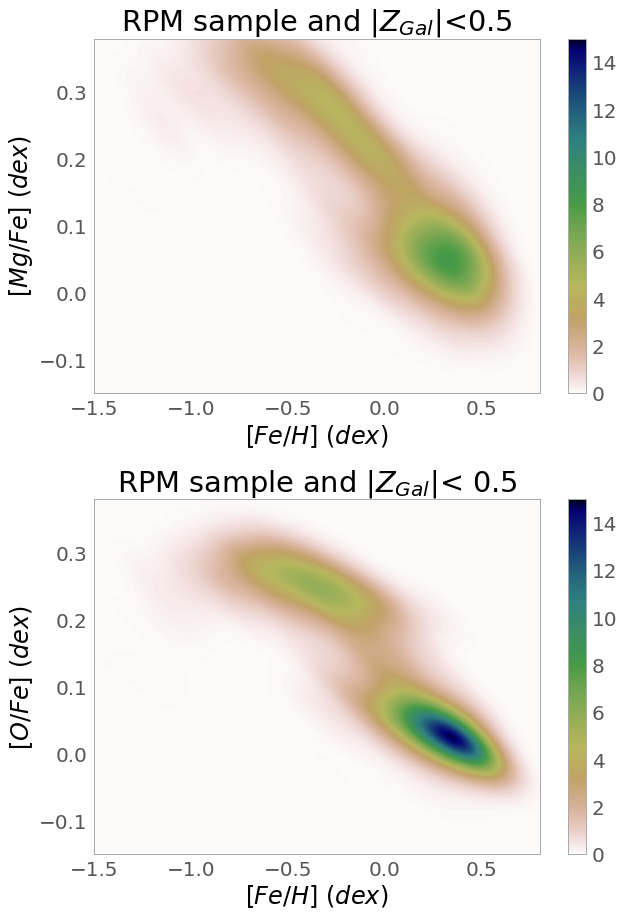}
\caption{ [O/Fe] vs. [Fe/H], and [Mg/Fe] vs. [Fe/H] for the RPM sample with an extra cut in |Z$_{Gal}$|<0.5 kpc, respectively. Here too, the figures are colour-coded according to probability density function.}
\label{fig:OMgfe}
\end{figure}

The metallicity distribution of our two samples is shown in Figure \ref{fig:mdf}. The Galactic bulge has long been reported to have multiple peak locations in the metallicity distribution \citep{McWilliam1997}, but the peak metallicity values vary considerably according to the sample and technique used (see Table 2 of \citealt{Barbuy2018}). From Figure \ref{fig:alphapop}, we select all the stars that fall within the highlighted red-dashed contour line in the upper left panel, and we plot the resulting metallicity distribution in Figure \ref{fig:mdf}. This region of stars has at least two peaks in the metallicity distribution: the most dominant peak at [Fe/H]= 0.30 and an intermediate peak at [Fe/H]=$-$0.68. This is in agreement with the peaks found by \cite{Rojas-Arriagada2020} $-$0.66, $-$0.17 and $+$0.32 dex, respectively. The multi-peaked metallicity distribution seen here can also be associated with different stellar populations in the Galactic bulge, as in \citet{Ness2013a}. However, there is no requirement for a physically motivated population to have a Gaussian or narrow chemical composition.
For a detailed study of the APOGEE DR16 MDF as a function of $(l,b)$ we refer to \citet{Rojas-Arriagada2020}. The MDF of our samples is discussed in Section \ref{orbital} in the context of the chemo-orbital analysis.


Finally, we also looked at two individual $\alpha$-elements, O and Mg, to ensure we obtain results that are  consistent with what is found using the $\alpha$ values obtained from the ASPCAP pipeline. Figure \ref{fig:omgfe} shows the [O/Fe] (with $13\,421$ stars) and [Mg/Fe] (with $13\,473$ stars) maps for the bulge--bar field sample. The results are consistent with the maps shown in Figure~\ref{fig:alphapop}. In Figure~\ref{fig:OMgfe}, which is similar to Figure~\ref{fig:alphafe} but made using [O/Fe] and [Mg/Fe] and only for the RPM sample, the  bimodality is still visible, though with a different morphology  when Mg is used. The different morphologies are most probably partly a consequence of the details of the stellar pipelines. The APOGEE/ASPCAP dispersion in uncertainties for [Mg/Fe] is higher for colder, low- to intermediate-metallicity stars \citep{Jonsson2020}.

\subsection{Checking for consistency with two other chemical clocks: [C/N] and [Mn/O]}
\label{cnmn}

\begin{figure}
\includegraphics[width=8.5cm]{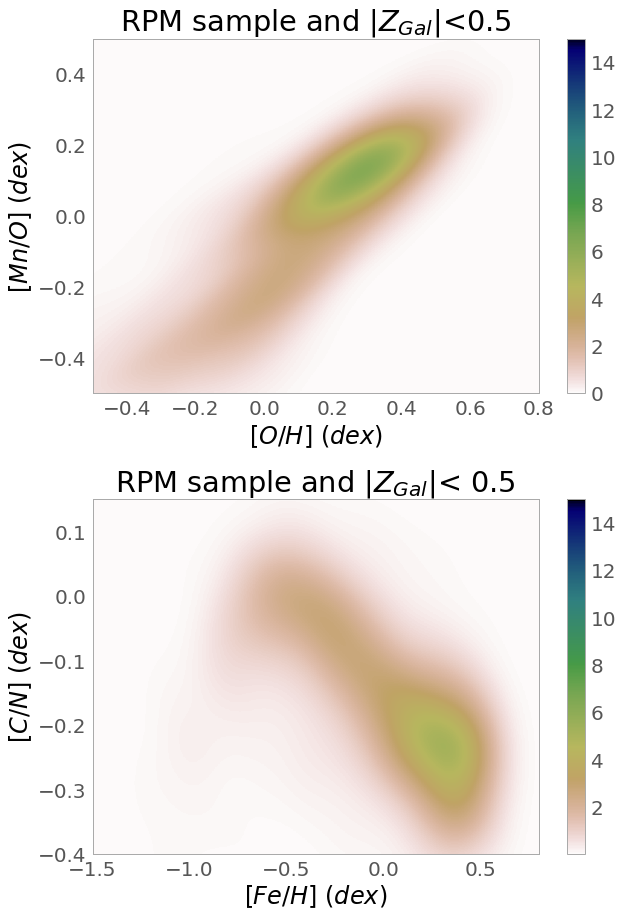}
\caption{Two other chemical clocks projected into 2D diagrams for the RPM sample at Z$_{Gal}$ $>$ 0.5 kpc. Upper panel:  [Mn/O] vs. [O/H]. Lower panel: [C/N] vs. [Fe/H]. Here too, the figures are colour-coded according to probability density function. }
\label{fig:MnCN}
\end{figure}


Other important chemical clocks are the [C/N] and [Mn/O] abundance ratios. The [C/N] is broadly dependent on stellar mass, because the first and third dredge-up  converts part of their C into N and thus decreasing the [C/N] ratio  \cite[see e.g.][]{Masseron2015}. The dependency of the [C/N] ratio at the solar vicinity has been shown to indicate a correlation with stellar ages coming from APOKASC \citep{Martig2016} for stars in the 7 $<$ R (kpc) $<$ 9 Galactocentric range. The usage of this ratio and its link to stellar age has been extrapolated to larger disc regions by \citet{Ness2016} and more recently by \citet{Hasselquist2019,Hasselquist2020}, although the dependency of the [C/N] ratio on metallicity in giants (both through hot bottom burning and stellar yields of C and N), and therefore on the chemical evolution of the Galaxy, makes these extrapolations very uncertain \cite[see][for a discussion]{Lagarde2019}.
Despite these caveats, the [C/N] map in Figure \ref{fig:cnmno} shows an encouraging agreement with previous maps based on the alpha elements, in the sense that larger [C/N] ratios correspond to high [$\alpha$/Fe] ratios, as expected.

The [Mn/O] ratio is also a very promising population tracer (see \citealt{Barbuy2018} for a discussion). This ratio should be low at earlier stages of chemical enrichment, when only core-collapse supernovae had time to pollute the ISM, increasing at later times due to the pollution by SNIa. However, its more complex nucleosynthesis \citep{Chiappini2003,Barbuy2013} makes this elemental ratio behave differently from other iron-peak ratios (especially, and most importantly,  at low metallicities), a fact that enhances differences between separate populations. An example is illustrated in Figure \ref{fig:cnmno}, where a nice correspondence between a low [Mn/O] ratio and the high [C/N] can again be observed. Nevertheless, our [Mn/Fe] distribution is biased against very cool stars, because the ASPCAP pipeline cannot properly measure Mn lines for stars with effective temperatures below 4000 K. This phenomenon is even more pronounced in the case of the RPM sample. Errors due to the assumption of local thermodynamic equilibrium (LTE) significantly affect data for Mn. \cite{Battistini2015}  showed that Mn trends can change drastically if non-LTE corrections are taken into account \cite[see also][]{Schultheis2017}.

The [Mn/O] and [C/N] ratios are projected in 2D diagrams in the panels of Figure \ref{fig:MnCN}. These panels still show hints of the bimodality observed in the $\alpha$-elements, despite their more complex nucleosynthesis, the lower statistical significance of these plots, and the larger uncertainties on the measurements of these abundance ratios from APOGEE spectra.


To summarise, in this section we confirm that the chemical bimodality previously observed in the alpha elements, is also present in the C/N and Mn/O ratios.
From the standpoint of bulge-formation chemodynamical models, the implications differ if one considers that the bimodality is formed by a continuous or two distinct star formation paths. 
The results presented here suggest a bimodality with a well-defined depression between the two peaks which is more in agreement with a discontinuous star formation path.

Different approaches, from pure chemical evolution to chemodynamical models (either isolated or in the cosmological scenario), have been explored to understand the observed chemical bimodality first seen around the solar vicinity, and more recently shown to extended towards the whole inner disc \citep{Queiroz2019} and bulge. These approaches are discussed in Section~\ref{concl}.

Finally, the chemical maps presented in this section show a consistent picture between the different tracers, and indicate the predominance of a moderately metal-poor \citep[][]{Barbuy2018,Savino2020} population in the innermost Galactocentric regions, which extends to larger Z$_{Gal}$. This population could be an extension  of the bulge RR Lyrae population ---discussed in the recent literature \citep{Kunder2020, Du2020}--- to more intermediate metallicities. We return to this discussion in Sect. \ref{concl}. Closer to the Galactic plane,  Z$_{Gal}<$300 pc, the metal-poor population is mixed with a much more metal-rich (and alpha-poor) population, which is very probably related to the rearrangement of disc stars forming a buckling bar. We now proceed to the analysis of the kinematical properties in this region.

\begin{figure*}
\centering
\includegraphics[width=16.5cm]{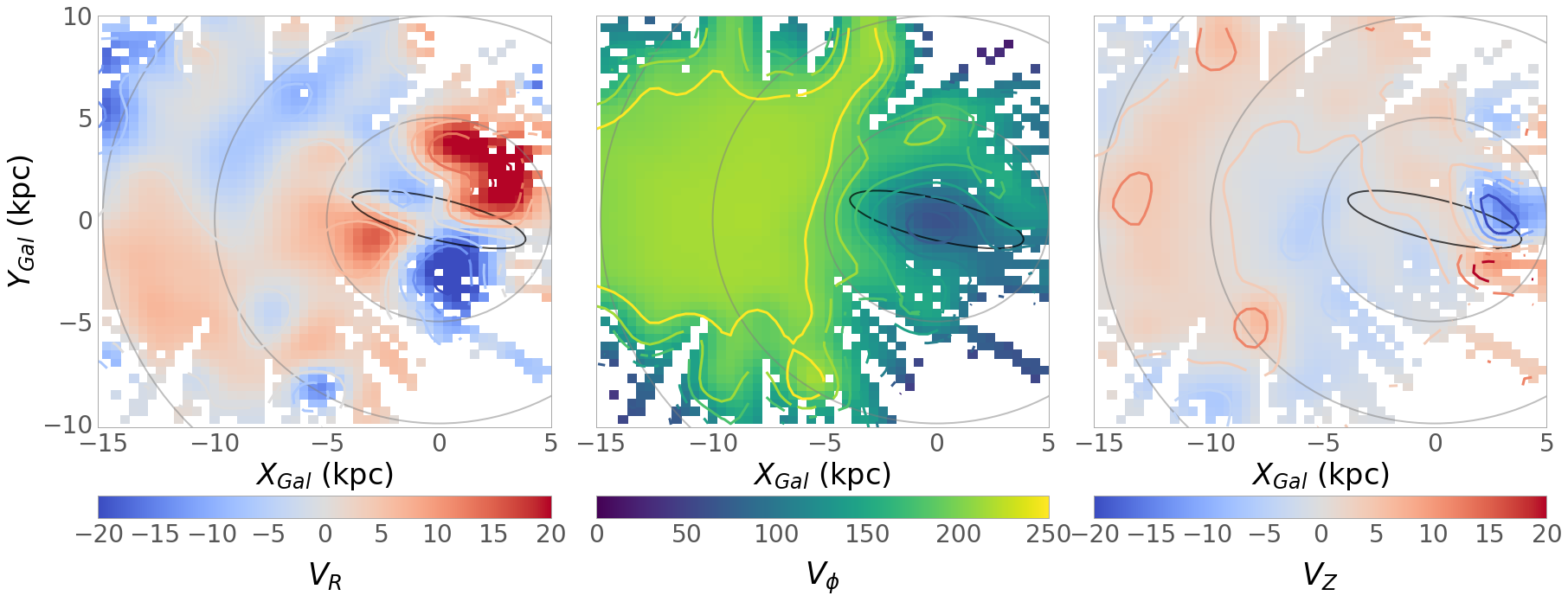}
\includegraphics[width=16.5cm]{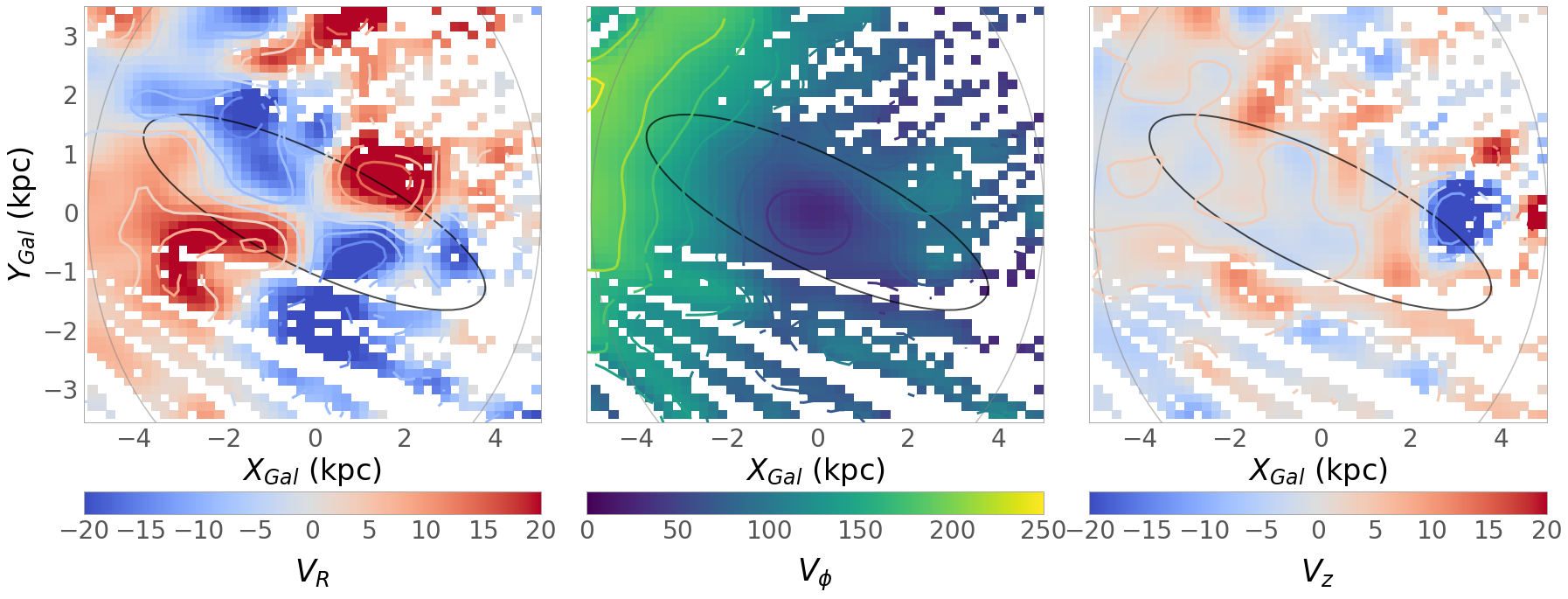}
\caption{Cartesian projection of the Galactic disc using {\tt StarHorse} distances. From left to right the maps are colour coded according to V$_{R}$ (first panel), V$_{\phi}$ (second panel), and V$_Z$ (last panel). Upper panels show the same region studied in \citet{Bovy2019}, while the lower panels show a zoom into the innermost 5 kpc of the Galaxy (the grey circles illustrate the Galactocentric distances of 5, 10, and 15 kpc. To guide the eye, in the figures we draw an ellipse with an inclination of 20 degrees in relation to the Sun--GC line, 4 kpc semi-major axis length, and 1 kpc semi-minor axis length.}
\label{fig:cylvel}
\end{figure*}

\section{Kinematics}\label{kine}

In Sect. \ref{chemo} we present the chemical-abundance distributions of our bulge--bar samples. The clear dichotomy between [$\alpha$/Fe]-rich/metal-poor and [$\alpha$/Fe]-poor/metal-rich stars suggests that the GC region is inhabited by (at least) two very distinct populations. In this section, we analyse the 3D velocity space to establish whether the two distinct chemical populations also present different kinematical properties.

\begin{figure*}
\centering
\includegraphics[width=15.5cm]{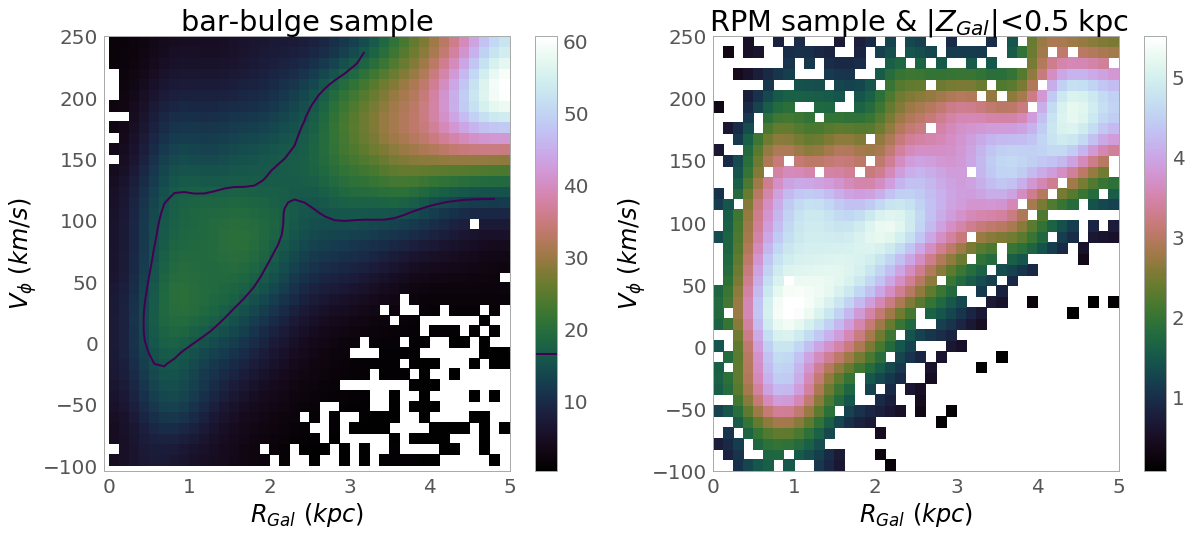}
\caption{  V$_{\phi}$ vs. Galactocentric radius for the inner  Galaxy, the entire bulge--bar sample (left panel), and the RPM with an additional cut in Galactocentric height (right panel). The left panel shows contours of density indicated by the colour bar, highlighting the kinematical populations present in the sample.}
\label{fig:vphir}
\end{figure*}

\begin{figure*}
\centering
\includegraphics[width=13.5cm]{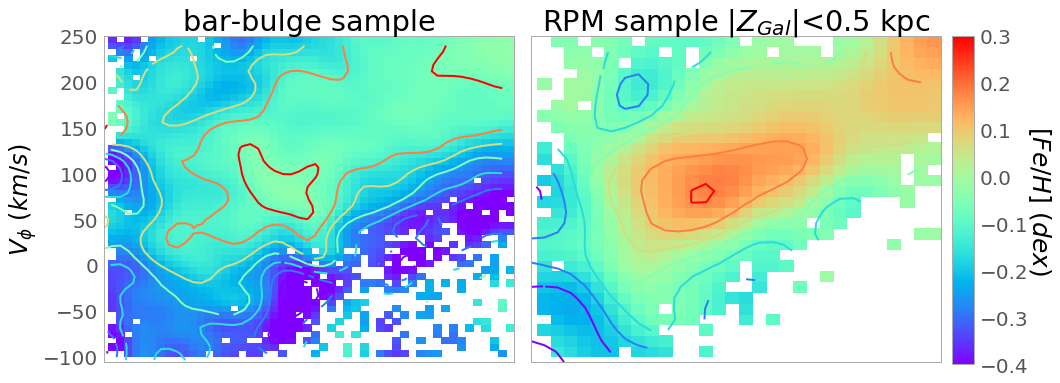}
\includegraphics[width=13.5cm]{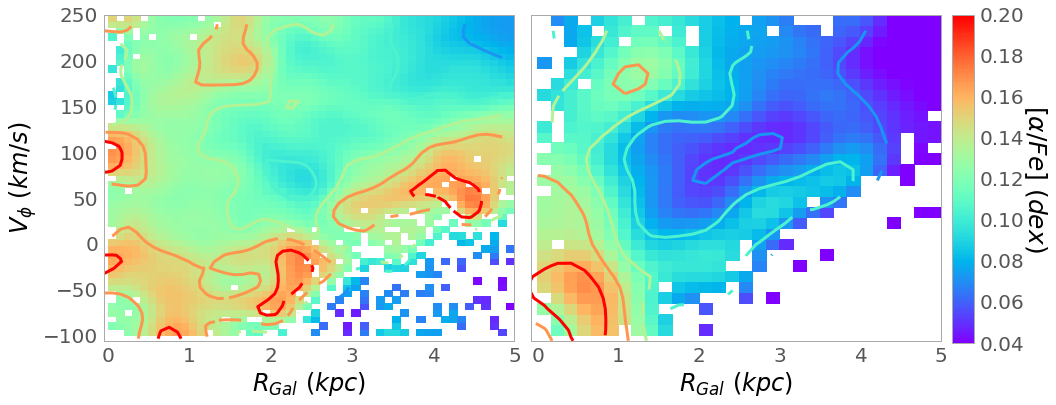}
\caption{Same as Figure \ref{fig:vphir}, but now with the panels colour-coded according to iron content (upper panels) and $\alpha$-elements (lower panels).}
\label{fig:vphiriron}
\end{figure*}

By combining {\it Gaia} EDR3 and APOGEE data, it has become possible to produce precise 3D kinematic maps that reach even the innermost parts of our Galaxy. 
\citet{Bovy2019} presented the first Cartesian maps of V$_\phi$ and V$_R$ using data from APOGEE DR16 coupled with distances obtained using the neural-network algorithm by \citet{Leung2019}. Figure~\ref{fig:cylvel}  shows X$_{Gal}$ versus Y$_{Gal}$ maps colour coded according to the three velocity components in the Galactocentric cylindrical frame. The maps in Figure~\ref{fig:cylvel} cover the bulge--bar sample with a cut in Z$_{Gal}<0.5$kpc (lower panels) and an extended region surrounding the Galactic disc (upper panels).

The signature of bar rotation is noticeable in Figure~\ref{fig:cylvel}. The first panel shows the Cartesian $X-Y$ map colour-coded according to $V_{R}$. A barred structure is expected to be characterised by a distribution of V$_R$ that extends both inward and outward along the bar. This is seen in simulations of barred galaxies, as discussed by \citet{Bovy2019} and \citet{Fragkoudi2020}. This effect is recognised in Figure~\ref{fig:cylvel} (first column, lower panel), where the resulting butterfly pattern of the $V_{R}$ field is clearly observed. A second and more extended quadrupole is seen in the upper panel of Figure~\ref{fig:cylvel}, indicating the presence of the spiral arms. By comparing the recent maps with simulations, it is possible to characterise the extent of the bar along both the major and minor axes, as well as its angle with Sun--GC line. A quantitative comparison with models is necessary to fully characterise the Galactic bar.

 The second panel of Figure~\ref{fig:cylvel}, colour-coded according to V$_{\phi}$, shows a more subtle elliptical shape extending in the $X_{Gal}$ axis by $\sim$ 2 kpc and in the $Y_{Gal}$ axis by $\sim$ 1 kpc, with the V$_{\phi}$ growing linearly from 0 to 150 km/s, which is a signature of the rigid body rotation of a barred structure. The elliptical structure in $V_\phi$ is not as extended and is also more spherical compared to \citet{Bovy2019}.

Finally, in the third panel of Figure~\ref{fig:cylvel}, we show V$_{Z}$. High positive V$_{Z}$ characterises the region situated on the right side of the ellipse. In contrast, an area with negative V$_{Z}$ is found at one end of the bar. In the extended velocity map (third column, upper panel) of Fig.~\ref{fig:cylvel}  positive V$_{Z}$ is seen in the outer disc, $\sim$ $10-12$ kpc, which was also reported by  \citet{Carrillo2019} based on {\it Gaia} DR2 and StarHorse data. The maps shown here show the wave structure in the disc much more clearly, extending the \citet{Carrillo2019} maps to a larger Galactocentric range.

In Figure~\ref{fig:vphir}, we plot the V$_{\phi}$ against Galactocentric radius for the bulge-bar sample and for the RPM selection with an extra cut in Galactic height ($|Z_{Gal}|<$ 0.5 kpc). These diagrams show the clear signature in the distinct stellar populations of a pressure-supported spheroid, a bar, and the Galactic discs. The first panel of Figure~\ref{fig:vphir} shows a population that has a high dispersion in V$_{\phi}$ within $R_{Gal}<$ 1 kpc and then a structure in which V$_{\phi}$ increases linearly with radius, and a third structure with V$_{\phi}$ of the order of that of the thin disc population, i.e. $\sim$ 200 km/s. When we apply the RPM cut (second panel of Figure~\ref{fig:vphir}), stars with  similar Galactic disc V$_{\phi}$ decrease significantly, indicating that our selection is indeed culling disc stars and leaving a purer bulge--bar sample. Biases must always be considered when analysing kinematics with a preceding selection in kinematics, but we would like to remind the reader that the cut in proper motions is subtle, and the velocity distributions of both samples do not change drastically apart from the clear decrease in stars at 200~km/s in $<V_{\phi}>$. The linear growth of $<V_{\phi}>$ with $R_{Gal}$ extends up to $\sim$ 4 kpc where there is a conglomeration of stars that could belong to  either the thick or the thin disc.

In order to confirm whether or not the kinematical structures seen in Figure \ref{fig:vphir} belong to different chemical populations,  in Figure \ref{fig:vphiriron} we reproduce  the same plot but colour-coded according to [Fe/H] and [$\alpha$/Fe]. High-metallicity, low-$\alpha$ stars are mostly concentrated around V$_{\phi} \sim 200$ km/s, which is again very consistent with what is expected for thin-disc stars. Metal-poor, [$\alpha$/Fe]-rich stars seem to be present in larger fractions inside R$_{\rm Gal} < 1$ kpc and to have a high V$_{\phi}$ dispersion, consistent with expectations for a pressure-supported spheroid. One may wonder from the figure what the two main concentrations of metal-poor stars are, one at negative V$_{\phi}$ and one around V$_{\phi}$ $\sim$ 100 km/s. This metal-poor V$_{\phi}$ bimodality in Figure~\ref{fig:vphiriron} is mainly caused by the large contribution of stars at V$_{\phi}$ $\sim$ 0, (see Figure~\ref{fig:vphir}). At V$_{\phi}$, R$_{Gal}$ $\sim$ 0, a more metal-rich and higher density component dominates, causing the bimodal metal-poor distribution. A bar population signature, where the V$_{\phi}$ grows linearly with radius, seems to be complex and characterised by a mixture of both metal-rich and metal-poor populations. However, it has a more considerable contribution from metal-rich stars, in agreement with the findings of \cite{Wegg2019}, but in contrast to those of \cite{Bovy2019} (see further discussion in Section \ref{zmaxecc}). A lump (blob) of high-metallicity stars is observed in the right panels of Figure \ref{fig:vphiriron}, between 10$<$ V$_\phi<$200 km/s and R$_{\rm Gal} \sim $3.5 kpc, which possibly represents the contribution of thin and thick disc stars in this region.
The maps in this section show the present position of the stars, which means that stars in halo or disc orbits could well be passing close to the GC and be confused with the inner stellar populations. With this in mind, we proceed to the orbital analysis of the RPM sample and its relation to chemical composition.

\section{Dissecting the mixed bulge populations in chemo-orbital parameters}
\label{orbital}

To further disentangle the mixed bulge populations that became evident during both the chemical (see Section \ref{chemo}) and kinematic analyses (see Section  \ref{kine}), we turn to an analysis of the 6D phase space distribution (for a description of the orbital parameters, see Section \ref{orbits}) and its relation to stellar chemistry.\\

\subsection{Counter-rotating stars}
\label{counter}

\begin{figure*}
\centering
\includegraphics[width=18.5cm]{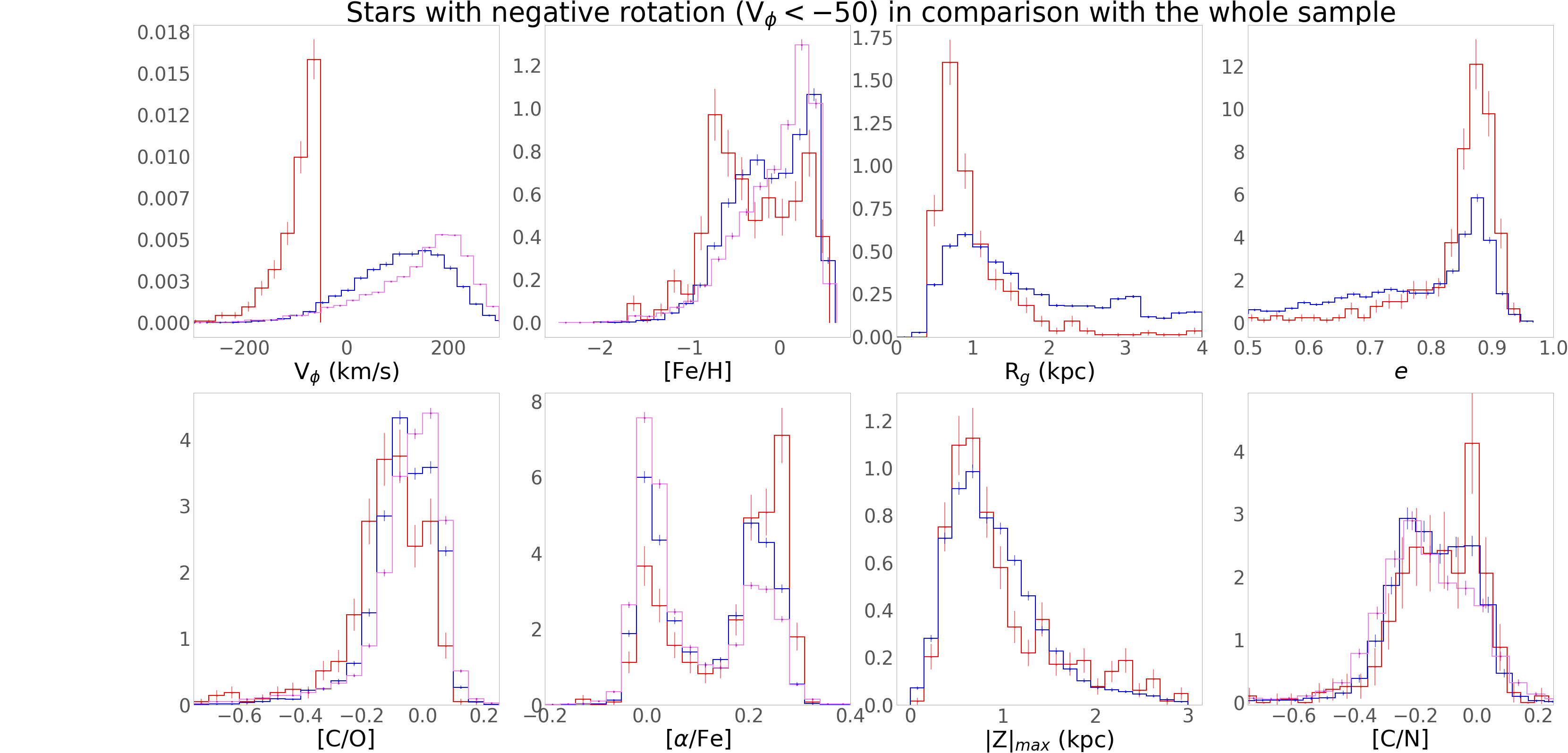}
\caption{Distribution of all stars with V$_\phi<$-50 km/s in the RPM sample shown in red as a function of many parameters, and compared to the same distribution for all stars in the RPM sample shown in dark blue, and all stars in the bulge--bar sample indicated by the violet lines.}
\label{fig:retro}
\end{figure*}

\begin{figure*}
\centering
\includegraphics[width=15.5cm]{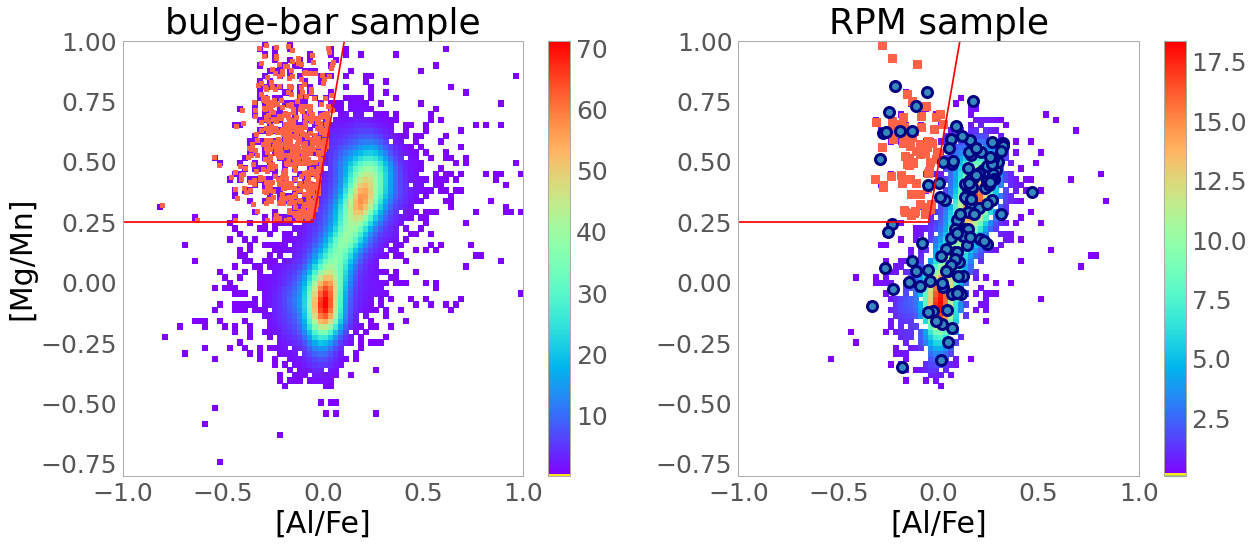}
\caption{[Mg/Mn] vs. [Al/Fe] diagram for the bulge--bar sample and the RPM sample. The red line indicates the locus of the accreted stars as defined by \citet{Hawkins2015} \citep[see also][]{Das2020}, and the stars that fall in this locus are indicated by the red squares. The blue dots represent the selection of counter-rotating stars, $V_\phi<-50$.}
\label{fig:mgmnal}
\end{figure*}

In Figure \ref{fig:vphiriron}, we notice a non-negligible contribution from stars with negative  V$_{\phi}$  that are mostly metal poor. We selected stars with V$_{\phi}<-$50 from the RPM sample, representing about 600 stars. In Appendix \ref{velflip}, we use Monte Carlo realisations to show that simple errors could not reproduce this tail of counter-rotating stars. In Figure \ref{fig:retro}, we analyse the properties of these stars.

\begin{figure*}
\centering
\includegraphics[width=16cm]{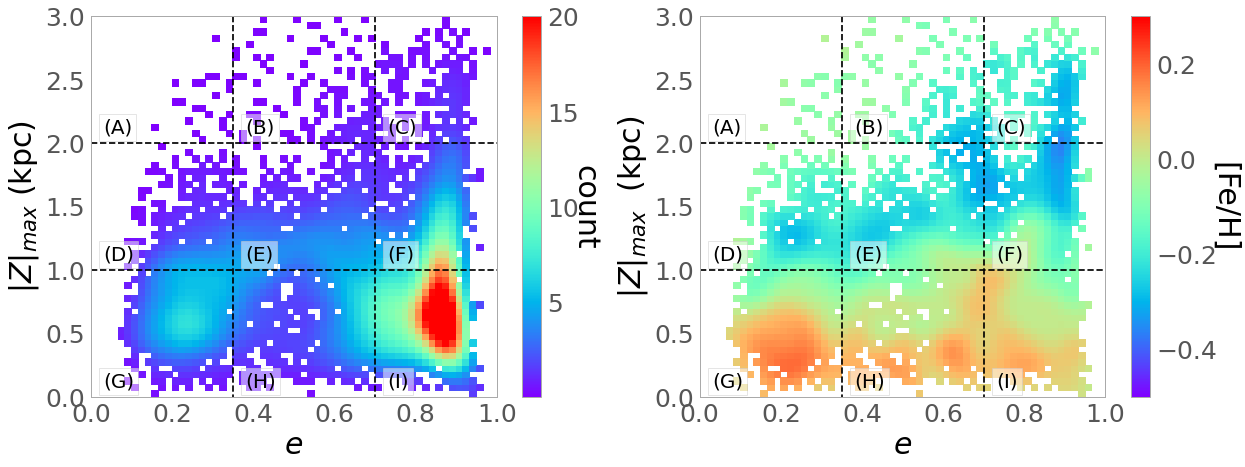}
\caption{|Z|$_{max}$ vs. eccentricity ($e$) diagram for the RPM sample. In the left panel, the colour shows the count of stars per bin and in the right panel the colour shows the mean [Fe/H] content per bin. We define nine windows in this diagram indicated by the letters (A) to (I).}
\label{fig:pmorbits}
\end{figure*}

Figure \ref{fig:retro} shows the distribution of parameters for stars in our RPM sample with V$_\phi<-$50 km/s in comparison with the full RPM and bulge--bar samples (limited always to Z$<0.5$kpc). The main properties of this retrograde population are as follows.

\begin{itemize}

\item Stars with V$_\phi<-$50 km/s are predominantly metal-poor, but show a broad metallicity distribution. The distribution has its highest peak at around [Fe/H] $\sim-0.7$, compatible with the metal-poor peak we see in Figure \ref{fig:mdf} at the inner GC.

\item The mean orbital radius distribution of the V$_\phi<-$50 km/s sample is confined to the innermost 1 kpc Galactocentric range, and the distribution has large eccentricities. 



\item Consistent with the fact that it is predominantly metal-poor, the retrograde population is [$\alpha$/Fe]-rich and [C/O]-poor (i.e. typical of gas mostly polluted by core-collapse supernovae).

\item The retrograde stars show larger [C/N] ratios, indicative of an older population (made of low-mass stars in which hot bottom burning does not take place, and therefore where C did not turn into N in these giants). 

\item Finally, we show an [Al/Fe] versus [Mg/Mn] diagram in Figure \ref{fig:mgmnal}. Our RPM sample automatically excludes most of the more obviously accreted population (in contrast to the sample selection of \citealt{Horta2020}). According to this criterion, the accreted stars are 500 out of $26\,500$ stars in the bulge--bar sample, and only 80 out of $8\,000$ stars in the RPM sample). The blue dots in the right panel of Figure \ref{fig:mgmnal} show the locus in the [Mg/Mn], [$\alpha$/Fe] plane of the counter-rotating stars. These blue dots are dispersed around the whole diagram and are not confined to the accreted location suggested by \citet{Hawkins2015}. As seen in the figure, we checked that the distribution of [Mg/Mn] for the retrograde component is shifted to larger values ($\sim$0.4), whereas a broader range in the values of [Al/Fe] is observed than that found for the accreted location defined by \citet{Hawkins2015}.
\end{itemize}

The origin of this highly eccentric and counter-rotating population confined to the innermost kpc of the Galaxy is unclear. One possibility is that this is an accreted metal-poor population originated during a gas-rich accretion phase in the early formation of the bulge. The metallicity distribution of the retrograde stars includes a metal-rich hump, but this could be explained by some contamination by metal-rich stars. Another interesting possibility is that we are seeing the inner Galaxy counterpart of the Splash population identified in the solar vicinity by \cite{Belokurov2020}. Splash stars have little to no angular momentum and many are on retrograde orbits and are slightly metal-poor, but can have a broad metallicity range. As explained by these latter authors, there are different theories for the origin of these stars, although the name Splash comes mainly from the idea that these are old stars that belonged to the proto-Galactic disc that were dispersed during the accretion event  that  created  the Gaia Sausage. However, alternative explanations are also possible. Among them are two very interesting notions that are more directly associated with bulge: (a) these stars were formed within gaseous outflows resulting from a burst in star formation or AGN activity \citep{Maiolino2017,Gallagher2019}, or (b) such retrograde stars in the bulge could be the result of clumps of star formation that took place at early times in the early disc (high redshift) and migrated into the bulge, with some stars being driven to retrograde orbits by the bar \citep{Amarante2020,Fiteni2021}. In both cases, it is expected that a broad velocity dispersion is created, with some stars being on counter-rotating orbits. A recent study analysing the kinematics of metal-poor stars in the inner Galaxy also found an extended tail of counter-rotating stars that does not match their simulations \citep{Lucey2021}.

Figures 16-19 illustrate the complexity of the Galactic bulge region. In addition to this counter-rotating hot component and/or tail, we see the contributions of other populations, with properties suggestive of a bar, an inner thin disc, a thick disc, and what seems to be a pressure-supported component that cannot be attributed to the halo or thick disc. As all these components overlap in the same region and parameter space, neither pure chemical nor kinematical criteria can be used to isolate these different populations. Therefore, we turn to a more detailed orbital--chemical analysis. Without pre-selections based on the classical definition of the local Galactic components, we can investigate the dominance of the different components around different parameter ranges.

\subsection{The $|Z|_{\rm max}$--eccentricity plane}
\label{zmaxecc}

We now turn to the analyses of our RPM sample in the $|Z|_{\rm max}$--eccentricity plane, similarly to that found in \cite{Boeche2013} and \cite{Steinmetz2020}. These latter studies showed that this parameter space offers a powerful way to disentangle the coexisting populations in the region (avoiding the use of pre-define Galactic populations based on properties of the more local samples). 

Figure \ref{fig:pmorbits} shows the distribution of stars in this plane colour-coded according to number density (left panel) and metallicity (right panel). We divide the $|Z|_{\rm max}$--eccentricity plane into nine cells (labelled alphabetically in the figure). From these diagrams we notice that most stars from our RPM sample have high eccentricity and low $|Z|_{\rm max}$.
A second prominent population is concentrated at very low eccentricities and low $|Z|_{\rm max}$, being mostly composed of high-metallicity stars, which is consistent  with classical disc populations. 
The right panel of Figure \ref{fig:pmorbits} is dominated by a metallicity gradient away from the midplane. On top of this, there is a population of less metal-rich stars on highly eccentric orbits that reaches  $\sim$ 1 kpc in $|Z|_{\rm max}$.  A deficit of stars is also noticeable at intermediate eccentricities of $\sim$ 0.48. Next, we analyse the composition distribution and orbital parameters for each cell.

\begin{figure*}
\centering
\includegraphics[width=15.5cm]
{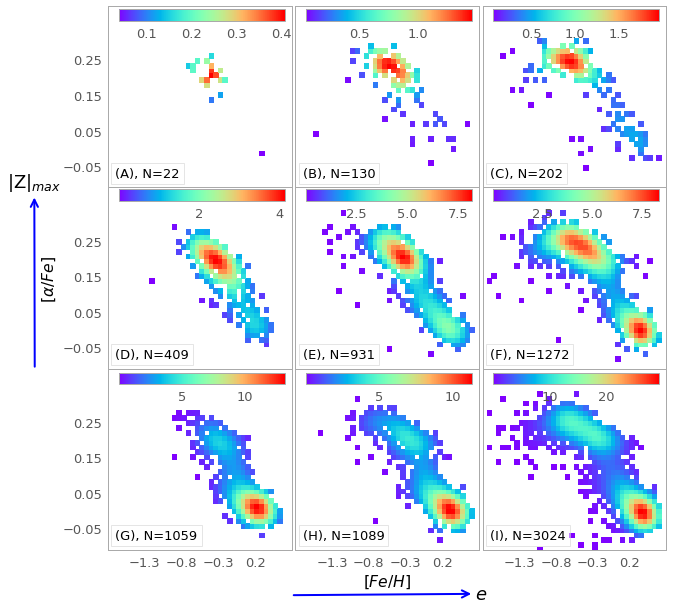}
\caption{[$\alpha$/Fe] vs. [Fe/H] for each cell defined in Fig. \ref{fig:pmorbits}. The number of stars for each  cell is indicated next to the panel labels. The rightmost columns, dominated by large eccentricity stars (pressure-supported component), show larger alpha-enhancement ([$\alpha$/Fe] $\sim$ 0.25) than what is seen among the low-eccentricity stars. The (inner) thin-disc contribution is seen mostly in the lower row, with a low, near-solar [$\alpha$/Fe] ratio, which peaks at [Fe/H] = 0.2 in  cells (G) and (H), and at 0.25 in cell (I).}
\label{fig:or_afe}
\end{figure*}

Figure \ref{fig:or_afe} shows [$\alpha$/Fe] versus [Fe/H] for each cell defined in Fig.~ \ref{fig:pmorbits}. We note that this is different from the usual diagram seen in bins of R, Z \citep[e.g.][]{Hayden2015,Queiroz2019}. Here instead we are focusing on a very inner sample, and mapping the chemistry of stars sampling different orbital parameter space in that inner region. This approach shows that low-[$\alpha$/Fe] stars are on low-inclination orbits, while high-alpha stars are on orbits of all types. Both populations are spread over orbits of every eccentricity.

Cell (I) shows a hot population (eccentricities $>$ 0.7) that is thin-disc-like and low-[$\alpha$/Fe] on top of a more metal-poor, high-alpha population. As we describe below, the stars in this cell are mostly stars on bar-shaped orbits.
As we go to higher $|Z|_{\rm max}$ we lose most of the low-[$\alpha$/Fe] stars, which results in the metallicity gradient seen in the right panels of Figure \ref{fig:pmorbits}. The separation between high-[$\alpha$/Fe] and low-[$\alpha$/Fe] is also clear in cell (I), whereas the bimodality becomes less clear for lower eccentricities and higher $|Z|_{\rm max}$.
 
The high-[$\alpha$/Fe] population shows a broad range of metallicities for the cells at high eccentricity (especially at low $|Z|_{\rm max}$) that gradually becomes narrower towards low eccentricities. The  cells (G) and (D) are consistent with predominantly thin and chemical-thick disc populations, respectively, with their distributions of $[\alpha/Fe] $ versus [Fe/H] appearing to be similar to those in \citet{Nidever2014,Hayden2015,Queiroz2019} for intermediate Galactocentric radii of 4 $<$ $R_{gal}$ $<$ 10 kpc. Note that when we refer to chemical-thick disc, we mean the definition of a thick disc by its high [$\alpha$/Fe] content. However, for stars on more eccentric orbits (cells C, F, and I), the high-[$\alpha$/Fe] populations become more extended in metallicity. One way of interpreting this is that the so-called knee moves to larger values for these stars. This is, for instance, the behaviour predicted for a spheroidal bulge (e.g. Matteucci et al. 2020, Cescutti et al. in prep). Moreover, these cells show slightly larger [$\alpha$/Fe] than those from the chemically defined thick disc in the solar neighbourhood. We note that this is not in contradiction with earlier APOGEE results showing that the high-[$\alpha$/Fe] chemical-thick-disc component has the same shape in different $R_{gal}$-$Z_{gal}$ bins; it is simply that now we are able to see a spheroidal population confined to the smallest  R$_{mean}$ that stands out among the more eccentric stars.
This suggests that the chemical-thick disc and spheroidal bulge have slightly different [$\alpha$/Fe]-enhancements \citep[see][for a discussion]{Barbuy2018}. 
We also should keep in mind that cells (G), (H), and (I) may be incomplete, because of the selection outside the heavily reddened regions as seen in Sect. \ref{sec:data}. 

\begin{figure*}
\centering
\includegraphics[width=15.5cm]{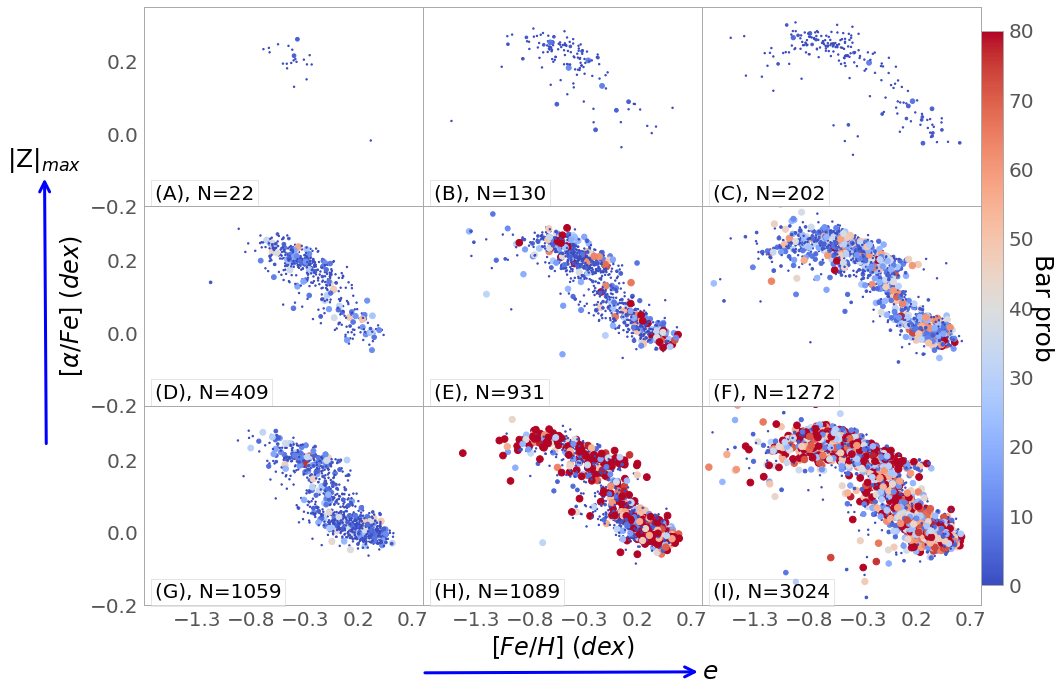}
\caption{[$\alpha$/Fe] vs. [Fe/H] for each cell defined in Figure \ref{fig:pmorbits} now colour-coded according to the probability that a star follows a bar-shaped orbit (see text). Stars with the largest bar-shaped orbit probabilities populate cells (H) and (I), and are found both among high- and low-[$\alpha$/Fe] populations.}
\label{fig:barprob}
\end{figure*}

To understand where bar-like orbits would fall in these diagrams, we made Figure~\ref{fig:barprob} which shows the [$\alpha$/Fe] versus [Fe/H] similarly to Figure \ref{fig:or_afe}, but now colour-coded according to the probability of the star moving on a bar-shaped orbit. To estimate this probability, we used the Monte Carlo sample of each star (50 orbits, see Sect. \ref{orbits}) and calculated the fraction of orbits classified as bar-shaped. To classify each orbit, we follow the definition from \citet{Portail2015} which uses frequency analysis. We compute the main frequencies of each orbit in the Cartesian coordinate x and the cylindrical radius R in the bar frame. The orbits for which the frequency ratio fR/fx = 2 +/- 0.1 are in a bar-shaped orbit. The orbits that are not bar-shaped have a frequency ratio fR/fx $\neq$ 2 +/- 0.1.

Figure \ref{fig:barprob} shows that the stars most likely to be on bar-shaped orbits are in cell (I), with an important contribution also found in cell (H). As expected, the stars on the bar show eccentric and low-$|Z|_{\rm max}$ orbits. 
One very important finding is that the stars following bar-shaped orbits in cells (I) and (H) are seen in both low- and high-$\alpha$ populations. This suggests that stellar trapping has been an efficient mechanism throughout the lifetime of the  bar, bringing stars to the bar that had previously belonged to Galactic components formed even before the bar was formed. There is a clear dearth of stars on bar-shaped orbits at high $|Z|_{\rm max}$ and with low eccentricity.

\begin{figure*}
\centering
\includegraphics[width=15.5cm]{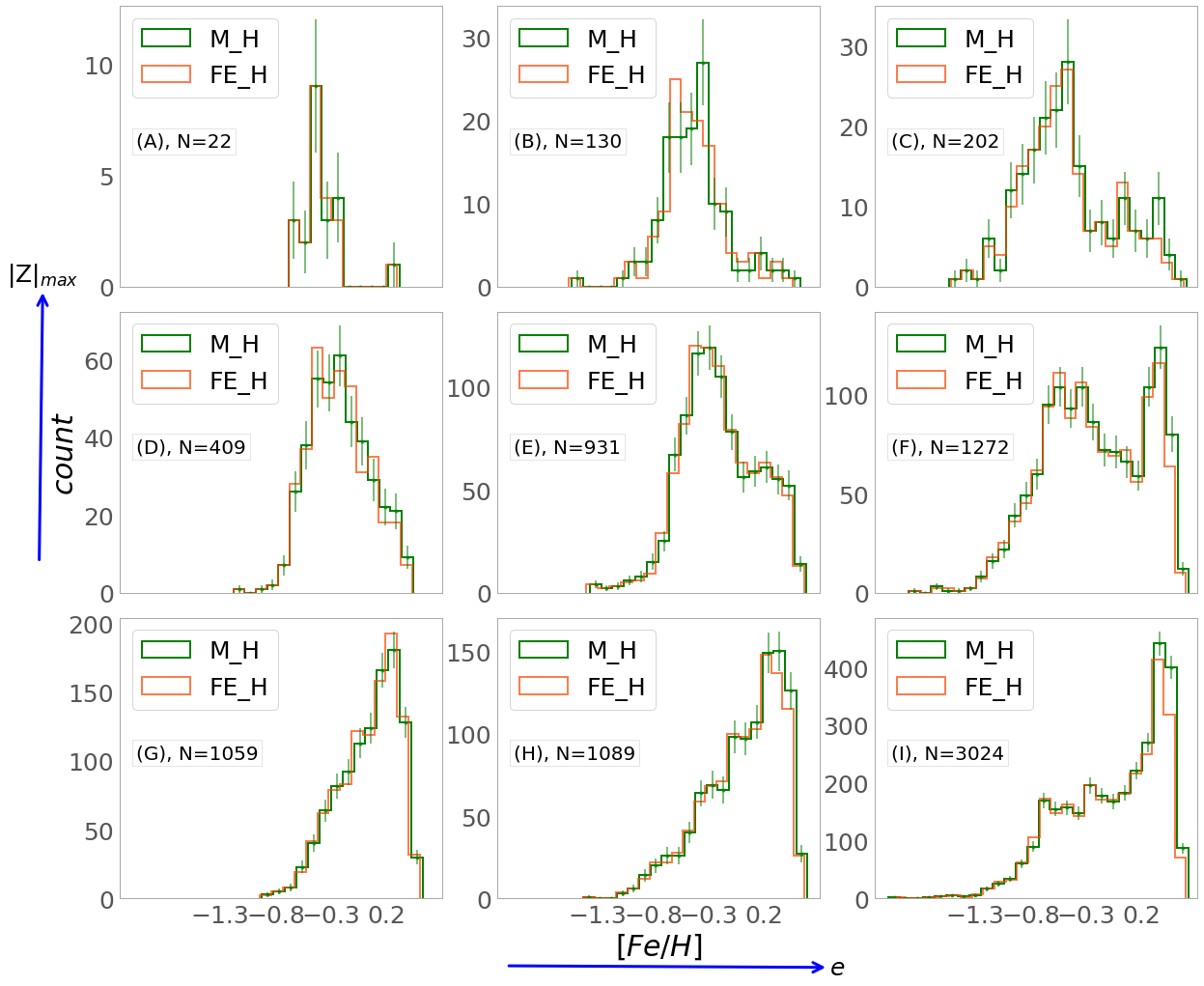}
\caption{MDF for each cell defined in Fig. \ref{fig:pmorbits}. We show the distributions of [Fe/H] (orange line) and [M/H] (green line) coming from ASPCAP. }
\label{fig:or_fe}
\end{figure*}

\begin{figure*}
\centering
\includegraphics[width=15.5cm]{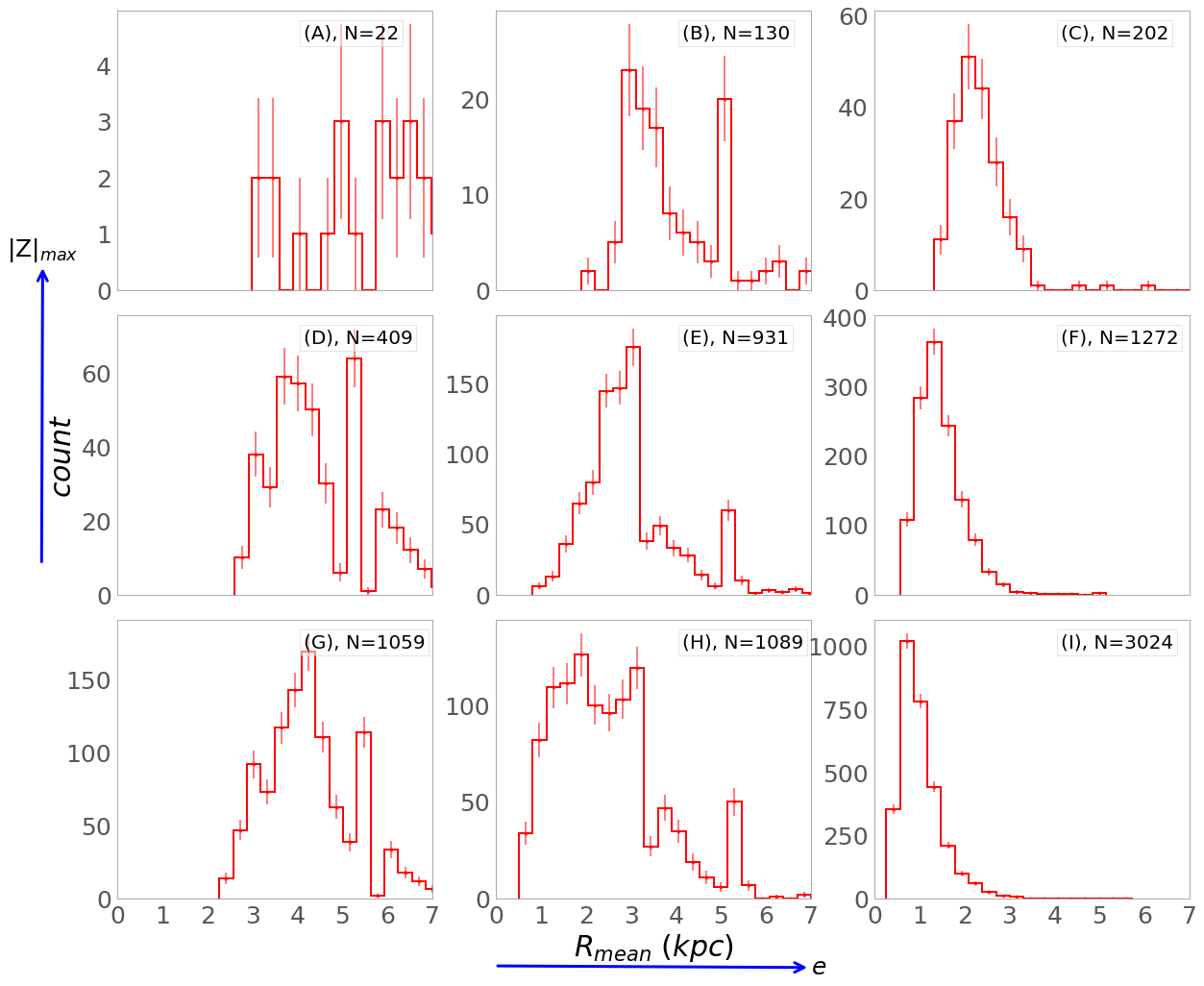}
\caption{Mean orbital radius distribution for each cell defined in Figure \ref{fig:pmorbits}. The more eccentric population has R$_{mean}$ confined to the innermost (1-3 kpc) regions of the Galaxy, whereas the thin-disc stars have R$_{mean}$ larger than 2-2.5 kpc.}
\label{fig:orRg}
\end{figure*}

Figures \ref{fig:or_fe} and \ref{fig:orRg} show the distributions of metallicity and  $R_{mean}$ for each $|Z|_{\rm max}$--eccentricity  cell. These figures show very interesting features that are related to what we see in the [$\alpha$/Fe]--[Fe/H] relationship discussed above.

In Fig. \ref{fig:or_fe} we see two populations, one with a narrow [Fe/H] centred on $\approx$ 0.2 and another, broader distribution centred on $\approx$ -0.7. Comparison of Figures \ref{fig:or_afe} and \ref{fig:or_fe} tell us that the former is the low-alpha population and the latter the high-alpha population. 
The high [$\alpha$/Fe] cells (I), (F), and (C) span the widest range of metallicities, but a narrower range in  R$_{mean}$, with most stars showing  R$_{mean}<$ 3 kpc. The sampled  R$_{mean}$ go from  R$_{mean}<$ 2 kpc (I) to 1 $<$ R$_{mean}<$ 3 kpc, as we go up in $|Z|_{\rm max}$. This is expected, but what is interesting is that this is accompanied by a low-metallicity component that starts to become  more prominent (going from cells I to C). As we show below, these high-eccentricity stars are composed of a mix of bar and spheroid populations, giving the impression of a metallicity gradient with $|Z|_{\rm max}$.
The peak in the metallicity of cell (C) is consistent with the metal-poor peak seen in Figure~\ref{fig:mdf}.
The metallicity distribution clearly becomes narrower towards lower eccentricities, while the distribution in R$_{mean}$ is now broader, and with fewer stars coming from the innermost kiloparsecs.
In the bottom row ($|Z|_{\rm max}$ < 1), the prominent high-metallicity peak goes from [Fe/H] $ \sim$ 0.25 in cell (I) to 0.2 in cells (G) and (H). Progressively, going from (I) to (G), the metal-poor population around $-$0.7 dex appears to get weaker (with fewer and fewer stars from the pressure-supported component, which is mostly composed of stars with R$_{mean}<$ 3 kpc). This is the population that is very dominant in cells (I), (F), and (C) as discussed before. Still in the bottom row, going from (I) to (G), a peak at [Fe/H] $ \sim -$0.3 gets more prominent. This peak will increase for intermediate eccentricities as $|Z|_{\rm max}$ < 1 increases. As seen here, this corresponds mostly to stars with 2 kpc  $<$  R$_{mean}<$ 3 kpc.

For low-eccentricity stars (left columns in Figures \ref{fig:or_fe} and \ref{fig:orRg}), the mean orbital radius distributions get broader, with R$_{mean}>$ 2 kpc. This suggests that the inner disc stars were not born in the innermost 2 kpc of the Galaxy, a result reminiscent of that of \citet{Matsunaga2016} based on classical Cepheids (see discussion in Section 7). The metallicity distribution is now dominated by stars in the 3 kpc  $<$ R$_{mean}<$ 4 kpc mean orbital radius range, and a peak around $-$0.27 dex starts to appear. In cell (G), the contribution of three peaks is visible at [Fe/H] $ \sim$ 0.2,$-$0.27, and $-$0.33 dex. Toward larger $|Z|_{\rm max}$ values, the metal-rich peak at $\sim$ 0.2 disappears, and the other two peaks begin to dominate, consistent with a transition from a thin-disc-like population to a thick disc population.

\begin{figure*}
\centering
\includegraphics[width=15.5cm]{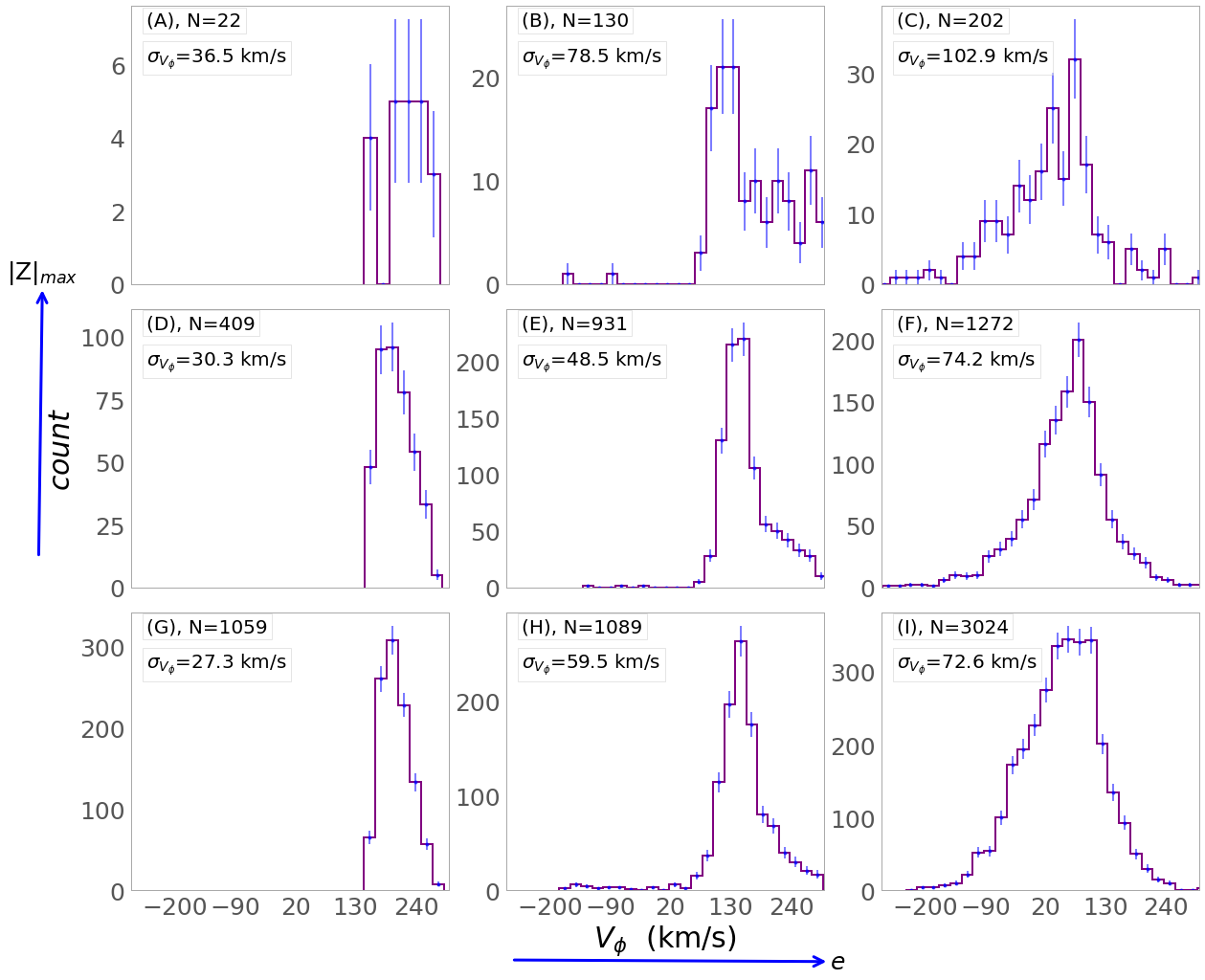}
\caption{V$_\phi$ distribution for each cell defined in Figure \ref{fig:pmorbits}. The value of the dispersion in V$_\phi$ is also shown for each cell. Thick disc stars with V$_\phi$ $\sim$ 140 km/s are seen in panels H and E where there is less contamination by thin disc stars (with V$_\phi$ around 200 k/ms - cells G and D) and the spheroidal component (with V$_\phi$ around 80 km/s cells I, F, and C). A counter-rotating tail is noticeable in cell C.}
\label{fig:orvphi}
\end{figure*}

By analysing Figures \ref{fig:or_fe} and \ref{fig:orRg}  together with the V$_\phi$ distributions (Figure \ref{fig:orvphi}), we can more quantitatively relate the populations discussed previously. It is possible to see the contributions from the inner thin and thick discs in Figures \ref{fig:orRg} and \ref{fig:orvphi} in cells (G),(D), and (A) (left column of the $|Z|_{\rm max}$--eccentricity diagram). The first column of the diagram is mostly dominated by inner thin-disc stars (stars with a V$_{\phi}$ peak at around 200 km/s and a low V$_{\phi}$ dispersion).
The second column of the $|Z|_{\rm max}$--eccentricity diagram (intermediate eccentricities) contains mostly thick disc-like stars, which become more dominant towards larger $|Z|_{\rm max}$ values (also confirmed by the metallicity distribution in Figure \ref{fig:or_fe}).
The last column of the $|Z|_{\rm max}$--eccentricity diagram (highly eccentric orbits) selects a pressure-supported component with lower rotation and larger V$_{\phi}$ dispersion (with small angular momentum and therefore small $R_g$ range),  which we saw in Fig.~\ref{fig:or_afe} to be a metal-poor, high-[$\alpha$/Fe] population. 

At low $|Z|_{\rm max}$ and high eccentricity (cell I), the bar population begins to dominate over the spheroid (pressure-supported population described in the previous paragraph), increasing the metallicity (as we also see in the bar probability figure).
The last column of Fig. \ref{fig:orvphi} also reveals, superposed on the spheroid and bar populations (both having large eccentricities), the counter-rotating, metal-poor population discussed in Section 7.1. Here, it is more prominent at the highest $|Z|_{\rm max}$  cell, probably because at lower $|Z|_{\rm max}$ it gets buried in the much more dominant metal-rich population of the bar. The counter-rotating population could also just be an extended tail of the spheroid. In Appendix \ref{velflip}, we show that the errors are not likely to form an asymmetric structure in V$_{\phi}$; significantly, that structure would extend to high negative rates such as -50 km/s.
We also notice positive tails in the three central panels of Figure \ref{fig:orvphi}. The canonical  V$_\phi$  distribution of an exponential disc has a sharp cutoff at high V$_\phi$, suggesting a slow outward decline in $\sigma_{R}$.

In summary, the analysis performed in this section shows, for the first time, a detailed dissection of the innermost parts of the Milky Way.
We show that the several peaks in the metallicity distribution correspond to populations of different eccentricities and $|Z|_{\rm max}$ distributions. The metal-rich population (with a peak at 0.2 dex) is made of inner thin disc stars, mostly formed outside the innermost 1-2 kpc. Some of the metal-rich stars are from the bar, which is populated by stars with  R$_{mean}$ within the 0-3 kpc range. These populations sit on top of a broader metallicity component extending from around $-0.8$ to above solar, which resembles a classical bulge \citep{Cescutti2018,Matteucci2019} made of mostly high-[$\alpha$/Fe] stars (most probably old; see \citealt{Miglio2021}). Meanwhile, with increasing $|Z|_{\rm max}$ we start to probe even more of the inner thick disc, and the metallicity distribution is increasingly dominated by stars with metallicities around $-$0.5 dex, which is similar to the peak of the local thick disc metallicity distribution (emerging in cell (B)).


\section{Summary and implications}
\label{concl}

In this paper, we analyse the inner regions of our Galaxy using APOGEE post-DR16 internal release data combined with {\it Gaia} EDR3 and the {\tt StarHorse} distances and extinctions. This latter addition provides us with an unprecedented catalogue of the Galactic innermost regions, with thousands of stars with distance uncertainties of less than 1 kpc. 

We analyse two distinct samples: (a) one sample of more than 26\,500 stars spatially selected in Cartesian coordinates X and Y, and (b) a sample of around  8\,000 stars that are more confined to the inner Galaxy and cleaned from foreground stars using the RPM method, which becomes possible thanks to the very precise proper motions of Gaia  EDR3. Most of this sample is outside the locus for accreted stars defined by \citet{Hawkins2015, Das2020} on the [Mg/Mn]--[Al/Fe] plane (but see discussion in \citealt{Horta2020}). Despite this, we see a counter-rotating population the origin of which requires further investigated (see discussion in Section 7.1).

With our larger sample, we were able to build exquisite chemical and kinematical maps of the innermost regions of the Galaxy. Furthermore, our analysis of the chemical data reveals a clear chemical bimodality in the [$\alpha$/Fe] versus [Fe/H] diagram for the full  sample of 26\,500 stars. The separation becomes more evident when we apply a proper-motion cut to clean the sample for foreground disc stars. Although the bimodality has also been detected in previous works \citep{Rojas-Arriagada2019,Queiroz2019} it is much more clearly seen here. We also confirm that similar results are obtained when we adopt [Mg/Fe] or [O/Fe]. This shows the level of precision and consistency obtained by the APOGEE ASPCAP pipeline \citep{GarciaPerez2016,Jonsson2020}.

In chemical evolution, it is expected that a bimodality seen in [$\alpha$/Fe] versus [Fe/H] is also seen in other chemical abundance ratios that trace similar enrichment timescales. Here we illustrate this using the [C/N] and [Mn/O] ratios. Indeed, double densities are also seen when they are plotted as a function of metallicity.
For the C/N ratio, the interpretation is complex as both elements can be modified during the evolution of the star on the giant branch. For [Mn/O], difficulties arise in the abundance measures because the  pipeline processing does not estimate Mn for stars cooler than 4000 K. Broadly the results remain consistent with the bimodality seen in alpha-elements.

The chemical maps show a spatial dependency on the metallicity, with the predominance of a metal-poor ($\alpha$-rich) component that is located in the central region. This feature can now be seen in the XY plane. This component is also seen on the [C/N] and [Mn/O] maps, again in agreement with nucleosynthetic sites of production of these different elements, and their release timescales to the interstellar medium. 

The  XY spatial maps of cylindrical velocities exhibit an elliptical but almost circular form in V$_{\phi}$ and a butterfly pattern in V$_{R}$, indicating the rotation of a barred structure, which is the kinematical signature of a bar. This is similar to what has been seen by \cite{Bovy2019}, also using DR16 data but with fewer stars and a completely different way of estimating distances. The velocity maps are in agreement with the expectation from simulations of barred galaxies, for example as discussed by several authors \citep{Debattista2017, Bovy2019, Carrillo2019, Fragkoudi2020}, where the butterfly pattern of the  $V_{R}$ field is one example of the expected features. These maps suggest an inclination of the bar with respect to the Sun--GC line of 20 degrees, and a spatial extent of around 4 kpc in the semi-major axis and 1 kpc in the semi-minor axis. A more detailed comparison with models is required to provide a more quantitative characterisation of the properties of the Milky Way bar.

 The V$_{\phi}$ versus Galactocentric radius (mapped both in [$\alpha$/Fe] and [Fe/H]) for the two samples studied here shows the signature of the distinct stellar populations coexisting in these samples, suggesting the presence of a pressure-supported spheroid, a bar, and the Galactic discs. These diagrams also show a counter-rotating population of metal-poor stars or an extended tail of negative V$_\phi$, which we then characterise in detail. In particular, the dispersion in  V$_{\phi}$ of the innermost metal-poor component is too large to be attributed to thick-disc stars (around 120 km/s), strongly suggesting the presence of an underlying spheroid, as predicted by \citet{Minniti1996}.

After the chemical and the velocity analysis we further dissect the innermost regions thanks to a sample of approximately 8\,000 stars for which we computed stellar orbits.
The populations are then characterised on a $|Z|_{\rm max}$--eccentricity plane (and in this way we avoid any pre-selection based on chemistry or kinematics).
 We pursue a joint analysis of the distributions of metallicity, [$\alpha$/Fe] abundance ratios, mean orbital radii (R$_{mean}$)\footnote{ R$_{mean}$ represents the mean Galactocentric distance a star has in its orbit, that is, the mean between its apocentric and pericentric distances. This is taken here as being close to the birthplace of the stars, except for effects due to radial migration.}, and the V$_\phi$ and its dispersion. This comprehensive analysis is needed in order to map the parameter space where each of the different populations dominates, thus avoiding the use of artificial sharp boundary definitions.
In this way, we identify and better characterise the chemical properties of the following populations inhabiting the innermost parts of the Milky Way close to the Galactic midplane:

\begin{itemize}

\item Inner thin-disc and the bar:

Most of the low-eccentricity, high-V$_{\phi}$ stars show low [$\alpha$/Fe].
This inner thin-disc population has a metallicity peak at [Fe/H] $= +$0.2. This metallicity shifts to larger values for more eccentric stars, still close to the Galactic midplane, reaching a peak of [Fe/H] $= +$0.25. However, these metallicities are seen only in the (1-2) kpc mean orbital radius range, suggesting that the most metal-rich stars are part of the bar component (in agreement with \citet{Wegg2019}). This suggests that the bar is slightly more enriched than the inner thin disc stars, most probably due to residual star formation in the innermost 2 kpc that form stars that enter bar orbits. Bars at high redshift could induce bursts of star formation due to gas trapping and gas funneling, especially toward the centre. The inner thin-disc stars in our sample have R$_{mean}$ larger than 3 kpc, consistent with the fact that the thin disc does not extend all the way to the GC (although this conclusion could be affected by the non-optimal coverage of the innermost regions). This is in agreement with a similar suggestion made by \cite{Matsunaga2016} who reported that no Cepheid was found in the innermost 2.5 kpc of the Milky Way.

\item Pressure-supported component and the bar: Underneath the bar population mostly found at large eccentricities and low heights from the plane (confirmed by the large fraction of stars with bar-shaped orbits in this part of the parameter space; see Figure~\ref{fig:barprob}), there is another component that is much broader in metallicity and that becomes more apparent towards larger distances from the Galactic midplane (where the bar component fades in). This large velocity dispersion component has a non-negligible contribution of metal-poor stars, which makes the metallicity distribution broad. This pressure-supported spheroid shows high-[$\alpha$/Fe] ratios.
Part of these spheroid stars that have orbits that are more confined to lower heights from the Galactic midplane also get trapped by the bar. Indeed, as shown in Figure~\ref{fig:barprob}, we find stars with a high probability of being in bar-shaped orbits among the high-[$\alpha$/Fe] stars.
This figure suggests that bar stars have eccentricities in the 0.5-0.8 range and metallicities above solar (explaining the shift to larger metallicities in cell (I) of Figure \ref{fig:or_fe}). 
Therefore, we find the bar to be composed mostly of metal-rich stars, with some additional contribution of stars with a similar chemical pattern to those in the spheroidal component. The latter were most probably trapped into the bar potential. It seems that the bar traps the more metal-rich component of the spheroid, while the more metal-poor component is able to escape the bar. The mechanisms that explain how this happens need to be investigated using proper dynamical models.
This also explains the details of the shape of the [$\alpha$/Fe] versus [Fe/H] distribution closer to the Galactic midplane, which becomes more metal rich both in high and low-alpha populations.

\item Inner chemical-thick disc: Stars of intermediate eccentricities with V$_\phi$ compatible with the local thick disc population dominate cell (E) (Fig.~\ref{fig:or_fe} and Fig.~\ref{fig:orvphi}). These stars show typical local thick disc metallicity distribution (Fig.~\ref{fig:or_fe}) and [$\alpha$/Fe] enhancement.
The majority of these stars are not on bar-shaped orbits.
Local thick discs stars were recently shown to be a very old and coeval population (thanks to the very precise ages from asteroseismology \citep[see][]{Miglio2021,Montalban2021}). Therefore, the same is expected to be true for the inner-thick disc population discussed here. If that is the case, it would suggest this component to have formed before the bar.

\item Counter-rotating stars or the tail of the pressure-supported (spheroid) component:
We find, superposed on the two components populating the high-eccentricity orbits (the bar and the pressure-supported spheroid), a population with negative V$_\phi$ in highly eccentric orbits, confined to the innermost kiloparsec of the Galaxy. 
This population is seen as a tail in the V$_{\phi}$ distribution shown in cell C of Figure~\ref{fig:orvphi}, and its properties are shown in Figure~\ref{fig:retro}. Given the low statistics of stars in cell C, a more robust characterisation of this population is deferred to future work when larger samples will be available.

\item The spheroid and the thick disc: The conclusion that we have a non-negligible contribution from a spheroid (on top of the thick-disc-like component)is strengthened by the shape of the high-alpha populations in Figure \ref{fig:or_afe}. The high-[$\alpha$/Fe] population can be seen to be shifted to slightly larger values of [$\alpha$/Fe] in the last column of Fig. 21 (spheroid-dominated) compared to the two other columns (more thick-disc dominated). Tthe extent of the high-alpha population is also different, going to larger metallicities for the spheroid-dominated population, suggesting a higher star formation rate (and efficiency) in the spheroidal bulge than in the thick disc. The caveat here is that this could also be the result of low statistics in the thick-disc-dominated cells. A more detailed comparison between these two populations, with more data, will be the topic of a forthcoming paper.

\end{itemize}

The existence of a spheroidal bulge in which star formation has been vigorous would be in agreement with what is expected from chemical evolution models (see a discussion in Section 4 of \citet{Barbuy2018}, and \citet{Matteucci2019}). In a scenario of fast enrichment, very old stars can be found already at metallicities [Fe/H] $\sim - $1 (see  \citet[][]{Chiappini2011,Wise2012,Cescutti2018} and Section 3.2.4 of \citet{Barbuy2018} for a discussion). Indeed, some of the oldest objects known in our Galaxy are located in the bulge and have metallicities around one-tenth of solar. For instance, the Galactic bulge has a system of globular clusters \citep{Minniti1995b} that are now known to be among the oldest in our Galaxy \citep{Barbuy2009, Chiappini2011,Barbuy2014,Kerber2018,Kerber2019,Ortolani2019}; these can be as old as the RR Lyrae. These stars were born around 400 000 years after the big bang, and are thus relics of the earliest chemical enrichment of the Universe.

The properties of the pressure-supported metal-poor, $\alpha$-enhanced stars we find in the bulge are consistent with the RR Lyrae stars in the same region. A debate over the origin of the RR Lyrae population in the bulge is ongoing, and the conclusions are very dependent on the samples analysed and models employed. Some of the  suggestions in the literature are that these RR Lyrae could be the extension of the stellar halo in the inner Galaxy \citep{Minniti1996,Perez-Villegas2017a}, have a bar distribution \citep{Pietrukowicz2015}, or show evidence of being a more spheroidal, concentrated, pressure-supported structure \citep{Dekany2013,Kunder2016,Contreras2018}. To break this dichotomy, \cite{Kunder2020} recently suggest the existence of two components of RR Lyrae in the inner Galaxy. One  RR Lyrae component is spatially and kinematically consistent with the bar, and the second component is more centrally concentrated and does not trace the bar structure. This agrees with the results shown here, where we see that the bar seems to trap mostly thin-disc stars, but also the more metal-rich part of the $\alpha$-enhanced spheroidal component.

The pressure-supported component could be the result of an accreted event or strong gas flows at the early stage of the formation of the  Galaxy, and this is consistent with an age for the RR Lyrae stars in the bulge of 13.41 $\pm$ 0.54 Gyr \citep{Savino2020}. \citet{Du2020} use  OGLE IV photometry and Gaia DR2 proper motions to analyse metal-poor ([Fe/H]$<-$1) and metal-rich ([Fe/H]$>-$1) RR Lyrae stars in the bulge. These authors concluded that the angular velocities and spatial distribution are different for metal-rich and metal-poor RR Lyrae stars. These results are in agreement with the findings of \citet{Wegg2019} and \citet{Kunder2020}.

The results presented here also offer some insight into the conundrum of the age of the bulge, namely: the
old ages from colour magnitude diagrams proper-motion-cleaned towards low extinction bulge windows versus the non-negligible contribution of stars younger than 5 Gyr suggested by the microlensed dwarfs \citep{Bensby2017}. After the analysis shown here, it is clear that each of the techniques leads to a different mixture of stars, with Baade's window CMD probing more of the spheroidal component mostly occurring in the inner 2-3 kpc of the Galaxy, whereas in the other case the stars are sampling a mix of spheroid and inner thin-disc stars, as confirmed by their multi-peak metallicity distribution \cite[see also][]{Rojas-Arriagada2020}.

The clear bimodality in the chemical diagrams for stars closer to the Galactic midplane and the existence of a dearth of stars in between the two overdensities (Fig.~\ref{fig:OMgfe}) offer an important new observational constraint to chemo-dynamic models of the Galaxy. There has been considerable debate over the origin of this bimodality based on data for stars closer to the solar vicinity, and since the proposition made more than 20 years ago by \citet{Chiappini1997} that this would reflect two main star formation paths, with a quenching of the star formation in between. More recently, this scenario has been revived both by chemical evolution models and numerical simulations \citep[e.g.][]{Anders2017a,Anders2018,Weinberg2019,Spitoni2021,Grand2020} as well as by the indication of an age dichotomy between the high- and low-$\alpha$ populations \citep{Miglio2021,Rendle2019,Lian2020,Das2020}. Cosmological simulations are particularly important to identify the reasons for this quenching, which can be manyfold, as discussed in the literature \citep[e.g.][]{Weinberg2019,Grand2020,Agertz2020,Ciuca2020,Buck2020,Vincenzo2020}. Alternative views, explaining the observed dichotomy as being the result of internal processes such as radial migration were also put forward \citep{Schonrich2009b,Sharma2020}, but difficulties in forming a hot thick-disc-like component by radial migration alone have been pointed out \citep[see][for a discussion]{Minchev2013,Minchev2016,Aumer2016a}. The data presented here for the innermost regions now show the dichotomy to also be present  in the innermost regions. The properties of the different populations show the dichotomy to be mainly a result of the mix of different populations.
The upper [$\alpha$/Fe] sequence is dominated by a spheroidal, pressure-supported component (the bulge) in the innermost 2-3 kpc, whereas it is dominated by thick disc stars beyond that distance. The lower sequence is formed by the bar in the innermost 2-3 kpc, and then by thin-disc stars not in the bar. Further out, the lower alpha-sequence is then the result of the thin disc mixture caused by radial migration from stars born at different Galactocentric distances \citep{Friedli1994,Minchev2013,Minchev2014}.
Stars born at different distances have different chemistry due to the inside-out formation of the disc. We note however that the chemical bimodality is less clear in the high-resolution data towards Baade's window (as can be seen in \citet{Barbuy2018}). However, in a recent study by \citet{Thorsbro2020} a chemical bimodality was also detected. Accurate distances are necessary to put these findings into a more robust context. One caveat we still have to consider is that even though {\tt StarHorse} provides a large improvement in distance and extinction estimates, it still does not take into account variations in the extinction law, which are potentially important in the bulge region. Improvements in this direction are also part of our future plans.

Finally, we also see a population of counter-rotating stars, which needs to be further investigated and confirmed. This population could be the remnant of an early accretion event, or the coalescence into the forming bulge of a clump of star formation formed by disc instabilities \citep{Elmegreen2008,Huertas2020} like those commonly observed in the discs of star-forming galaxies at redshift z $\sim$ 2-3. Otherwise it could simply be the tail of the large dispersion spheroid.

APOGEE plus Gaia have been transformative in our understanding of the innermost parts of the Milky Way. 
The picture emerging from our results is in much better agreement with high-redshift observations, which show early spheroids being formed due to massive amounts of highly dissipative gas accretion and mergers as suggested by simulations (e.g. \citealt{Tacchella2015,Bournaud2016,Renzini2018}).

\section*{Acknowledgements}
The authors thank the referee, Prof. James Binney, for all the valuable suggestions.
The authors thank R. Schiavon for helpful discussions. 
CC acknowledges support from DFG Grant CH1188/2-1 and from the ChETEC COST Action (CA16117), supported by COST (European Cooperation in Science and Technology).
FA is grateful for funding from the European Union's Horizon 2020 research and innovation programme under the Marie Sk\l{}odowska-Curie grant agreement No. 800502 H2020-MSCA-IF-EF-2017. 
BB acknowledges partial financial support from FAPESP, CNPq, and CAPES - Finance Code 001. 
APV acknowledges the FAPESP postdoctoral fellowship no. 2017/15893-1 and the DGAPA-PAPIIT grant IG100319. 
DAGH acknowledges support from the State Research Agency (AEI) of the Spanish Ministry of Science, Innovation and Universities (MCIU) and the European Regional Development Fund (FEDER) under grant AYA2017-88254-P. 
ARA acknowledges partial support from FONDECYT through grant 3180203. J.G.F-T is supported by FONDECYT No. 3180210 and Becas Iberoam\'erica Investigador 2019, Banco Santander Chile. S.H. is supported by an NSF Astronomy and Astrophysics Postdoctoral Fellowship under award AST-1801940.
SH is supported by an NSF Astronomy and Astrophysics Postdoctoral Fellowship under award AST-1801940.
ABAQ, CC, FA, BX, BB acknowledges support from Laboratório Interinstitucional de e-Astronomia (LIneA).

This work has made use of data from the European Space Agency (ESA)
mission {\it Gaia} (\url{http://www.cosmos.esa.int/gaia}), processed by the {\it Gaia} Data Processing and Analysis Consortium (DPAC,
\url{http://www.cosmos.esa.int/web/gaia/dpac/consortium}). Funding
for the DPAC has been provided by national institutions, in particular the institutions participating in the {\it Gaia} Multilateral Agreement. \\

The {\tt StarHorse} code is written in python 3.6 and makes use of several community-developed python packages, among them {\tt astropy} \citep{AstropyCollaboration2013}, {\tt ezpadova}, {\tt numpy} and {\tt scipy} \citep{Virtanen2019}, and {\tt matplotlib} \citep{Hunter2007}. The code also makes use of the photometric filter database of VOSA \citep{Bayo2008}, developed under the Spanish Virtual Observatory project supported from the Spanish MICINN through grant AyA2011-24052.\\

Funding for the SDSS Brazilian Participation Group has been provided by the Minist\'erio de Ci\^encia e Tecnologia (MCT), Funda\c{c}\~ao Carlos Chagas Filho de Amparo \`a Pesquisa do Estado do Rio de Janeiro (FAPERJ), Conselho Nacional de Desenvolvimento Cient\'{\i}fico e Tecnol\'ogico (CNPq), and Financiadora de Estudos e Projetos (FINEP).\\

Funding for the Sloan Digital Sky Survey IV has been provided by the Alfred P. Sloan Foundation, the U.S. Department of Energy Office of Science, and the Participating Institutions. SDSS-IV acknowledges support and resources from the Center for High-Performance Computing at the University of Utah. The SDSS web site is \url{www.sdss.org}.\\

SDSS-IV is managed by the Astrophysical Research Consortium for the Participating Institutions of the SDSS Collaboration including the Brazilian Participation Group, the Carnegie Institution for Science, Carnegie Mellon University, the Chilean Participation Group, the French Participation Group, Harvard-Smithsonian Center for Astrophysics, Instituto de Astrof\'isica de Canarias, The Johns Hopkins University, 
Kavli Institute for the Physics and Mathematics of the Universe (IPMU) / University of Tokyo, Lawrence Berkeley National Laboratory, 
Leibniz-Institut f\"ur Astrophysik Potsdam (AIP),  
Max-Planck-Institut f\"ur Astronomie (MPIA Heidelberg), 
Max-Planck-Institut f\"ur Astrophysik (MPA Garching), 
Max-Planck-Institut f\"ur Extraterrestrische Physik (MPE), 
National Astronomical Observatory of China, New Mexico State University, 
New York University, University of Notre Dame, 
Observat\'ario Nacional / MCTI, The Ohio State University, 
Pennsylvania State University, Shanghai Astronomical Observatory, 
United Kingdom Participation Group,
Universidad Nacional Aut\'onoma de M\'exico, Univity of Arizona, 
University of Colorado Boulder, University of Oxford, University of Portsmouth, 
University of Utah, University of Virginia, University of Washington, University of Wisconsin, 
Vanderbilt University, and Yale University.\



\bibliographystyle{aa}
\bibliography{bar_bulge_apogee_aa-final}


\appendix

\section{Probabilities of flipping the velocity}\label{velflip}

Here we decipher whether or not the errors in velocity would produce inconsistent results in our analysis, especially the case where errors can cause the measured parameter to flip its original sign. This situation could fabricate the counter-rotating bump we observe in Section \ref{counter}.

\begin{figure*}{}
\centering
\includegraphics[width=18cm]{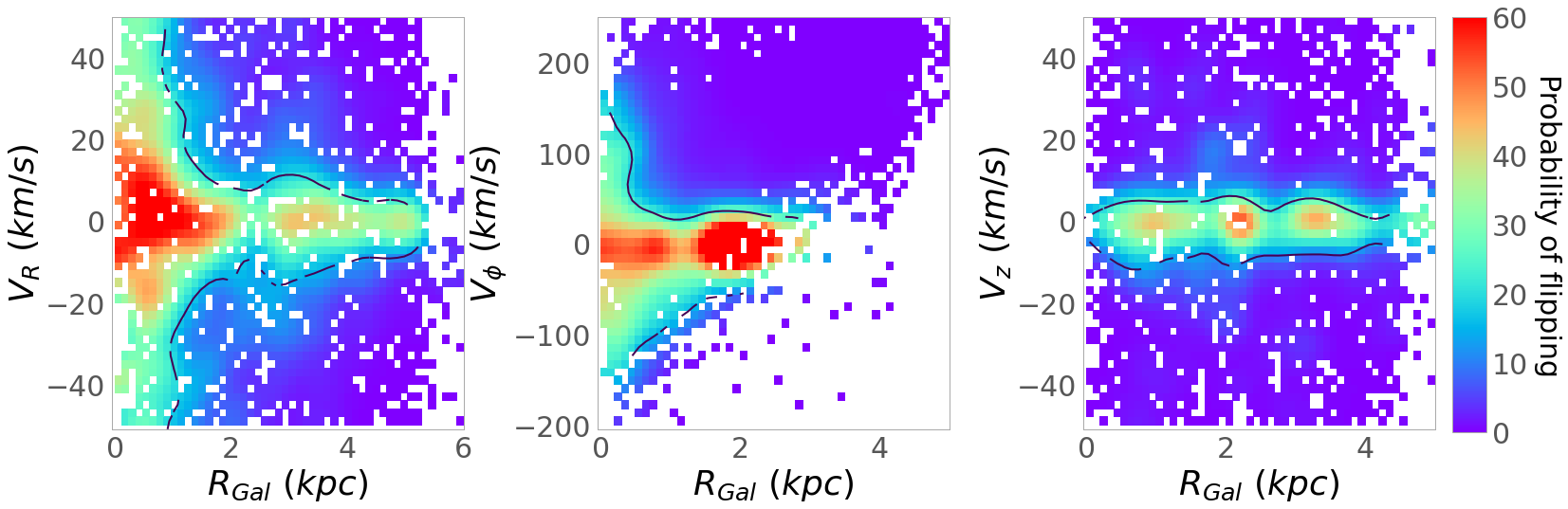}
\caption{ From left to right: Radial, azimuthal, and vertical velocities against Galactocentric radius. The diagrams are colour coded according to the  probability that velocity will flip sign.}
\label{fig:flipv}
\end{figure*}

To prove this is not the case in our data, we performed 1000 Monte Carlo realisations, considering the errors in the distance,  the line-of-sight velocity, and the proper motions to calculate the probability of the star flipping its velocity. The parameter that most influences the error in velocities is the distance. Figure \ref{fig:flipv} shows the median velocity component against Galactocentric distance colour coded according to the probability of flipping the sign. As can be seen in the figure, this probability is higher for small velocities, in the case of $V_\phi$ < -50 km/s. For $V_\phi$ we have that 61\% of the stars in the RPM sample will never change sign; from the $\sim 8\,000$ stars, $\sim 1\,000$ have more than 50\% probability of changing direction. If we split the stars with $>$ 50\% of changing sign in positive and negative, we have 559 $(\sim7\%)$ that go from positive to negative and $458 (\sim6\%)$ that go from positive to negative. This shows that the errors in $V_\phi$ are symmetric and would not likely produce the extended tail in velocities $<-50 km/s$ we see in Section \ref{counter}. The flipping probabilities are also symmetric in the other components of the velocity, as one can see from Figure \ref{fig:flipv}.

\section{Orbits comparison}
\begin{figure*}{}
\centering
\includegraphics[width=18cm]{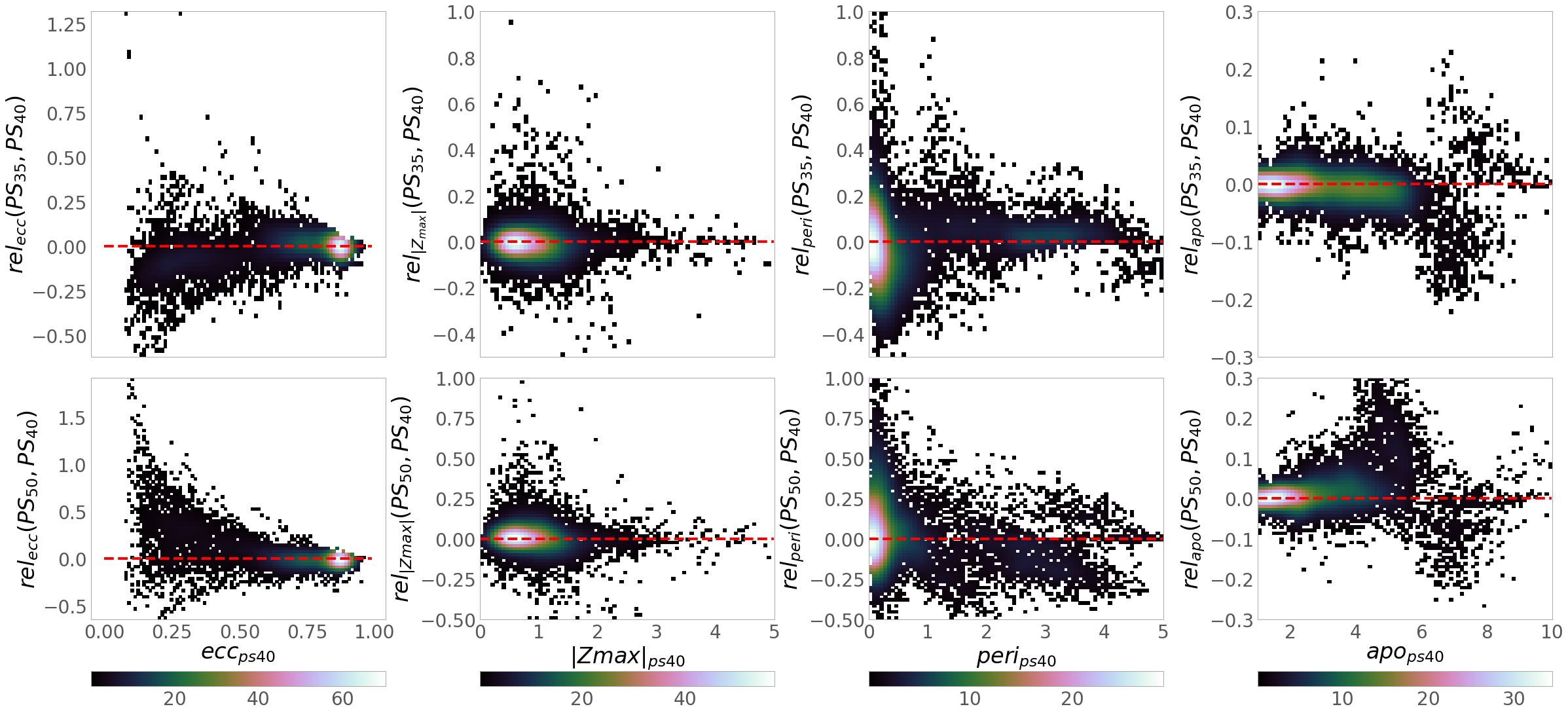}
\caption{Comparison between orbits with the same bar potential but with varying pattern speeds. From left to right relative errors for eccentricity, Z$_{max}$, pericentre, apocentre. The upper panels compare pattern speeds of 40 and 35; lower panels pattern speeds 40 and 50.}
\label{fig:orbtsps}
\end{figure*}

Here we show the differences in the orbital parameters if they were calculated with different pattern speeds for the bar potential. We verified that using a different pattern speed does not lead to any inconsistency in the presented results; Figure \ref{fig:orbtsps} shows the relative errors between eccentricity, $|Z|_{max}$, pericentre, and apocentre for two different pattern speeds of 35 and 50. The relative errors are generally not higher than 25\%. Errors are more significant for low-eccentricity stars and pericenter determination.

\end{document}